\documentclass[10pt,english,aps,prd,nofootinbib,onecolumn,floatfix]{revtex4-1}
\usepackage{CJK}

\usepackage{amsfonts}
\usepackage{amsmath}
\usepackage{hyperref}
\usepackage{amssymb}
\usepackage{mathtools}
\usepackage{cleveref}
\usepackage{cancel}
\usepackage{xfrac}
\usepackage{xcolor}
\usepackage{relsize}

\usepackage{tikz}
\usepackage[com pat=1.1.0]{tikz-feynman}
\usetikzlibrary{calc,arrows,decorations.markings,shapes}
\usepackage{pgfplots}
\pgfplotsset{compat=1.3}
\usetikzlibrary{3d,calc}

\tikzset{every picture/.style={
    scale=0.5, transform shape,
    }}

\usepackage{units}

\usepackage{xspace}

\usepackage{color}
\usepackage{graphicx}
\usepackage[space]{grffile}
\usepackage{verbatim}
\usepackage{amsmath}
\usepackage{amssymb}
\usepackage{wasysym}
\usepackage{ifsym}%
\usepackage{url}
\usepackage{bbold}
\usepackage{slashed}
\usepackage{epstopdf}
\usepackage{braket}
\usepackage{float}

\usepackage{support-caption}
\usepackage{subcaption}

\newcommand{\boundellipse}[3]
{[black,fill=blue!30] (#1) ellipse (#2 and #3)
}
\newcommand{\boundellipseW}[3]
{[white,fill=white] (#1) ellipse (#2 and #3)
}

\newcounter{countitems}
\newcounter{nextitemizecount}
\newcommand{\setupcountitems}{%
  \stepcounter{nextitemizecount}%
  \setcounter{countitems}{0}%
  \preto\item{\stepcounter{countitems}}%
}
\makeatletter
\newcommand{\computecountitems}{%
  \edef\@currentlabel{\number\c@countitems}%
  \label{countitems@\number\numexpr\value{nextitemizecount}-1\relax}%
}
\newcommand{\nextitemizecount}{%
  \getrefnumber{countitems@\number\c@nextitemizecount}%
}
\newcommand{\previtemizecount}{%
  \getrefnumber{countitems@\number\numexpr\value{nextitemizecount}-1\relax}%
}
\makeatother    
{\end{itemize}%
\unskip\computecountitems\ifnumcomp{\previtemizecount}{>}{3}{\end{multicols}}{}}

\begin{document}

\title{A gauged non-universal \texorpdfstring{$U(1)_{X}$}{} model to study muon \texorpdfstring{$g-2$}{} and $B$ meson anomalies}

\author{J. S. Alvarado$^{1}$}
\email{jsalvaradog@unal.edu.co}
\author{S. F. Mantilla$^{2}$}
\email{mantilla@pks.mpg.de}
\author{R. Martinez$^{1}$}
\email{remartinezm@unal.edu.co}
\author{F. Ochoa$^{1}$}
\email{faochoap@unal.edu.co}
\author{Cristian Sierra$^{3}$}
\email{cristian.sierra@njnu.edu.cn}

\affiliation{$^1$Departamento de Física, Universidad Nacional de Colombia, Ciudad Universitaria, K. 45 No. 26-85, Bogotá D.C., Colombia}

\affiliation{$^2$Max-Planck Institute for the Physics of Complex Systems, D-01187 Dresden, Germany}


\affiliation{$^3$Department of Physics and Institute of Theoretical Physics, Nanjing Normal University, Nanjing, Jiangsu 210023, China}

\date{\today}

\begin{abstract}
We study a non-universal $U(1)_{X}$ extension of the Standard Model with an extended scalar sector of two doublets and one singlet plus three additional exotic quarks and two exotic charged leptons on the fermionic sector. In order to obtain the observed fermion mass hierarchy, an additional $\mathbb{Z}_{2}$ discrete symmetry is imposed, where the heaviest fermions acquire their masses from three different scales determined by two Higgs doublets and one singlet, whereas the lightest fermions obtain their masses from effective operators up to dimension seven. From chiral anomalies cancellation, the model also includes heavy right-handed neutrinos which get massive via an inverse see-saw mechanism, reproducing the observed mass differences for the active neutrinos. We analyze phenomenological consequences of the model in view of the so-called flavour anomalies, namely the latest measurement made by Fermilab of the anomalous magnetic moment of the muon $g-2$, and additionally, the fit to semi-leptonic $B$ meson decays made by different flavour groups, dominated mainly by the LHCb 2020 data. We obtain that the model can explain the former at the $1\sigma$ level by means of contributions coming from charged $W^{+}$ bosons interacting with exotic Majorana neutrinos at one-loop level, with the $Z'$ boson contribution itself coming from the $U(1)_X$ symmetry being negligible. However, we find that the model is able to accommodate the $b\to s\ell^{+}\ell^{-}$ transitions associated to the $B$ meson anomalies only at the $2\sigma$ level via tree-level $Z'$ boson exchange, while simultaneously respecting various constraints from the recent $R_{K^{(*)}}$ measurements made by the LHCb, neutrino trident production and $B-\bar B$ oscillations.

\end{abstract}

\maketitle	

\section{Introduction}

Several extensions of the Standard Model (SM) have arisen in an attempt to explain the so called flavour anomalies, with the largest deviations being the anomalous magnetic moment of the muon $g-2$, and the semi-leptonic quark level decays involving $b\to s$ transitions related to many branching ratios and angular observables of $B$ meson decays. Regarding $g-2$, the Fermilab Muon $g-2$ Experiment, based on data collected in 2019 and 2020, has recently reported the most precise measurement of $a_{\mu}=(g-2)/2$ (improving the previous result \cite{Muong-2:2021ojo} by more than a factor of two) in combination with the Brookhaven National Laboratory (BNL)\cite{FermiLab:2023,g-2exp1,*g-2exp2} 
\begin{align}
    a_{\mu}^\textrm{Exp}=116592059(22)\times 10^{-11}\,(0.19\textrm{ ppm}),
\end{align}

\noindent value which is larger than the SM prediction $a_{\mu}^{\textrm{SM}} = 116591810(43)\times10^{-11}$ reported by the the Muon $g-2$ Theory Initiative's White Paper \cite{Aoyama:2020ynm} by
\begin{equation} \label{eqn:gm2discrepancy}
    \Delta a_{\mu} = 249\pm48\times10^{-11},  
\end{equation}

\noindent representing a discrepancy of $5.1\sigma$ between the SM and the experimental values\footnote{This deviation has to be taken with care given the discrepancy on the SM prediction from different theoretical working groups.}. The uncertainty of the experimental value is expected to decrease over time, with more data to be published by the Fermilab experiment and improving the precision by a factor of three \cite{FermiLab:2023} by 2025, and also the J-PARC experiment \cite{J-PARC} to provide an independent measurement. \\

With respect to the $B$ meson anomalies, and from a new physics (NP) perspective\footnote{For a recent discussion on the hadronic effects interpretation see \cite{Ciuchini:2022wbq}.}, the most recent global fits prefer a vector-like lepton universal (LU) coupling with a pull of around $6\sigma$ with respect to the SM \cite{Alguero:2023jeh}, including the latest LHCb measurements for $R_{K^{(*)}}$ \cite{LHCb:2022qnv,LHCb:2022vje},

\begin{equation}
\begin{split}
    0.1<q^2<1.1 \begin{cases}
        R_K &= 0.994~^{+0.090}_{-0.082} (\mathrm{stat}) ^{+0.029}_{-0.027} (\mathrm{syst}), \\
        R_{K^{\ast}} &= 0.927~^{+0.093}_{-0.087} (\mathrm{stat}) ^{+0.036}_{-0.035}(\mathrm{syst}),
   \end{cases} \\
   1.1<q^2<6.0
   \begin{cases}
        R_K &= 0.949~^{+0.042}_{-0.041} (\mathrm{stat}) ^{+0.022}_{-0.022} (\mathrm{syst}), \\
        R_{K^{\ast}} &=1.027~^{+0.072}_{-0.068} (\mathrm{stat}) ^{+0.027}_{-0.026} (\mathrm{syst}),
   \end{cases}
\end{split}
\end{equation}

\noindent now in agreement with the SM at $0.2$ standard deviations and not affecting significantly the preference for a vector-like LU coupling in the global fits for the one-dimensional scenarios \cite{Greljo:2022jac,Alguero:2023jeh}. Other significant flavour anomalies are the $B$ meson decays related to $b\to c \ell\nu_{\ell}$ transitions, the Cabibbo angle anomaly (CAA), leptonic $\tau$ decays of the form $\tau\to\mu\nu\nu$ and non-resonant di-electrons \cite{Crivellin:2022qcj} which we will not address here. \\

Among the most sounding NP explanations for those anomalies, two Higgs doublet models (2HDMs) \cite{2HDM1,*2HDM2,*2HDM3,*2HDM4,*2HDM5,*2HDM6,*2HDM7,*2HDM8,*2HDM9}, gauged $L_{\mu}-L_{\tau}$ \cite{mu-tau,*mu-tau1,mu-tau2}, leptoquarks \cite{model6,*model6-1,*model6-2,*model6-3,*model6-4,*model6-5,*model6-6,*model6-7,*model6-8,*model6-9,*model6-10,*model6-11,*model6-12,*model6-13,*model6-14,*model6-15,*model6-16,*model6-17,*model6-18,*model6-19,*model6-20,*model6-21,*model6-22,*model6-23,*model6-24,*model6-25,*model6-26,*model6-27} and  $Z'$ boson models \cite{model5,*model5-1,*model5-2,*model5-3,*model5-4,*model5-5,*model5-6} stand out.  In this paper, we will focus on the latter type of models, which are well motivated and have been extensively studied in the literature (for a review see e.g., \cite{Langacker2008yv}). Specifically, we will consider the gauged non-universal $U(1)_X$ extension presented in \cite{orig}. In this model, and as a result of requiring a theory free of chiral anomalies \cite{Ellis:2017nrp,Allanach:2018vjg}, heavy exotic fermions are introduced in the particle spectrum, adding then extra degrees of freedom which, along the $Z^{\prime}$ boson, could in principle explain both the anomalous $g-2$ and the deviations on $B$ meson semileptonic observables. \\

A compelling feature of the proposed $U(1)_X$ model is that it can explain the mass hierarchy of both quarks and charged leptons as well as neutrino oscillations throughout an inverse see-saw mechanism \cite{orig}. The lightest fermions, i.e., the down and strange quarks, and the electron obtain their masses radiatively \cite{Garnica:2019hvn}. However, in the present work, we show that those masses for the lightest fermions can also be generated by the Froggatt-Nielsen mechanism \cite{Froggatt:1978nt}. Another interesting characteristic of the model is that the imposed $\mathbb{Z}_2$ symmetry could be replaced by a Peccei-Quinn symmetry in order to generate the same required mass matrix textures while simultaneously addressing the strong CP problem \cite{Garnica:2019hvn}. \\

This paper is structured as follows: First we give an overview of the proposed Abelian extension in section \ref{modelgen}, then masses and rotation matrices for leptons and quarks are introduced in section \ref{fermions}, where effective operators are considered in order to give masses to the lightest fermions. Later, we compute the contributions at leading order (LO) of the new particle spectrum to muon $g-2$ in section \ref{g-2anom} and we illustrate how the $U(1)_X$ model can explain the $g-2$ anomaly at the $1\sigma$ level by means of one-loop contributions involving the SM $W$ boson and three heavy TeV Majorana neutrinos. After this, we calculate the largest tree-level contribution mediated in this case by the $Z^{\prime}$ boson to the $B$ meson observables in section \ref{Banom} via effective Wilson coefficients, finding a viable parameter region at the $2\sigma$ level. Finally, we summarize our conclusions in section \ref{conclusions} and present calculations for the boson and fermion masses in the appendix.

\section{The \texorpdfstring{$U(1)_{X}$}{U(1)X} extension}\label{modelgen}

\begin{table}[ht]
    \centering
    \begin{tabular}{|ccc|ccc|} \hline \hline
        Scalar Doublets & & & Scalar Singlets & &  \\
         & $X^{\pm}$ & $Y$ & & $X^{\pm}$ & $Y$   \\ \hline \hline
    $\small{\phi_{1}=\begin{pmatrix}\phi_{1}^{+}\\\frac{h_{1}+v_{1}+i\eta_{1}}{\sqrt{2}}\end{pmatrix}}$ & $\sfrac{+2}{3}^{+}$ & $+1$ & $\chi=\frac{\xi_{\chi}+v_{\chi}+i\zeta_{\chi}}{\sqrt{2}}$	&	$\sfrac{-1}{3}^{+}$	&	$0$	\\
    $\small{\phi_{2}=\begin{pmatrix}\phi_{2}^{+}\\\frac{h_{2}+v_{2}+i\eta_{2}}{\sqrt{2}}\end{pmatrix}}$&$\sfrac{+1}{3}^{-}$&$+1$&  & &  \\\hline
    \end{tabular}
\caption{Scalar particle content of the model, $X$ charge, $\mathbb{Z}_{2}$ parity ($\pm$) and hypercharge $Y$.}
\label{scalarlist}
\end{table}

The scalar particle spectrum of the Abelian extension is presented in table \ref{scalarlist}. It consists of two scalar doublets $\phi_{1,2}$ with vacuum expectation values $\upsilon_{1,2}$ such that $\upsilon=\sqrt{\upsilon_1^2+\upsilon_2^2}$ for $\upsilon=246\,\mathrm{GeV}$, and one scalar singlet $\chi$ associated to the spontaneous symmetry breaking (SSB) of the $U(1)_{X}$ symmetry through its respective VEV, $v_{\chi}$, expected to be at the TeV scale and endowing with mass the associated $Z'$ gauge boson as shown in appendices \ref{scalars} and \ref{gaugebosons}.  \\

The $U(1)_X$ symmetry assigns  non-universal $X$ charges to the SM particle content, that along the $\mathbb{Z}_{2}$ symmetry, can generate textures for the mass fermion matrices suitable for explaining the observed fermion mass hierarchy. As a result of this charge assignation, exotic heavy fermions at the $v_{\chi}$ scale are introduced among the SM ones from the requirement of having a theory free of the following triangle anomaly equations,

\begin{eqnarray}
\label{eq:Chiral-anomalies}
\left[\mathrm{\mathrm{SU}(3)}_{C} \right]^{2} \mathrm{\mathrm{U}(1)}_{X} \rightarrow & A_{C} &= \sum_{Q}\left[X_{Q_{L}} - X_{Q_{R}} \right],	  \nonumber\\
\left[\mathrm{\mathrm{SU}(2)}_{L} \right]^{2} \mathrm{\mathrm{U}(1)}_{X} \rightarrow & A_{L}  &= \sum_{\ell}\left[X_{\ell_{L}} + 3X_{Q_{L}} \right],	\nonumber	\\
\left[\mathrm{\mathrm{U}(1)}_{Y} \right]^{2}   \mathrm{\mathrm{U}(1)}_{X} \rightarrow & A_{Y^{2}}&=
	\sum_{\ell, Q}\left[Y_{\ell_{L}}^{2}X_{\ell_{L}}+3Y_{Q_{L}}^{2}X_{Q_{L}} \right] - \sum_{\ell,Q}\left[Y_{\ell_{R}}^{2}X_{L_{R}}+3Y_{Q_{R}}^{2}X_{Q_{R}} \right],	\nonumber	\\
\mathrm{\mathrm{U}(1)}_{Y}   \left[\mathrm{\mathrm{U}(1)}_{X} \right]^{2} \rightarrow & A_{Y}&=
	\sum_{\ell, Q}\left[Y_{\ell_{L}}X_{\ell_{L}}^{2}+3Y_{Q_{L}}X_{Q_{L}}^{2} \right]- \sum_{\ell, Q}\left[Y_{\ell_{R}}X_{\ell_{R}}^{2}+3Y_{Q_{R}}X_{Q_{R}}^{2} \right],	\nonumber	\\
\left[\mathrm{\mathrm{U}(1)}_{X} \right]^{3} \rightarrow & A_{X}&=
	\sum_{\ell, Q}\left[X_{\ell_{L}}^{3}+3X_{Q_{L}}^{3} 	-X_{\ell_{R}}^{3}-3X_{Q_{R}}^{3} \right] 	\nonumber,	\\	
\left[\mathrm{Grav} \right]^{2}   \mathrm{\mathrm{U}(1)}_{X} \rightarrow & A_{\mathrm{G}}&=
	\sum_{\ell, Q}\left[X_{\ell_{L}}+3X_{Q_{L}} -X_{\ell_{R}}-3X_{Q_{R}} \right], \label{eq:chiral_anom}
\end{eqnarray}

\noindent where the second equation counts only $SU(2)$ doublets, $Q$ and $\ell$ runs over all quarks and leptons respectively and $Y$ is the corresponding weak hypercharge. The electric charge definition given by the Gell-Mann-Nishijima relationship remains unaltered as $Q=\sigma_{3}/2+Y/2$ with $\sigma^{a}$ the Pauli matrices. \\

By scanning values for the charges $X=\pm2/3,\,\pm1/3$ for quarks and $X=\pm1,\,0$ for leptons and taking into account the scalar content in table \ref{scalarlist} along the introduced $\mathbb{Z}_2$ symmetry, we found the particular solution shown in table \ref{fermionlist} of the chiral anomaly equations (\ref{eq:chiral_anom}). The resulting quark sector considers one exotic up-like quark $\mathcal{T}$ while there are two additional down-like particles $\mathcal{J}^{1,2}$. On the leptonic sector, there are two exotic charged lepton singlets $E$ and $\mathcal{E}$ and three right-handed neutrinos $\nu_{R}$. In addition to those and without affecting the chiral anomalies, three Majorana neutrinos $\mathcal{N}_{R}$ are also introduced, being necessary to provide appropriate masses to the active neutrinos via an inverse see-saw mechanism \cite{orig}.

\begin{table}[ht]
\centering
\begin{tabular}{|cccc|ccc|}
\hline\hline
Quarks	&	$X$	&$\mathbb{Z}_{2}$&&	Leptons	&	$X$&$\mathbb{Z}_{2}$	\\ \hline 
$q^{1}_{L}=\left(\begin{array}{c}u^{1} \\ d^{1} \end{array}\right)_{L}$
	&	$+1/3$	&$+$	&&
$\ell^{e}_{L}=\left(\begin{array}{c}\nu^{e} \\ e^{e} \end{array}\right)_{L}$
	&	$0$	&$+$	\\
$q^{2}_{L}=\left(\begin{array}{c}u^{2} \\ d^{2} \end{array}\right)_{L}$
	&	$0$	&$-$	&&
$\ell^{\mu}_{L}=\left(\begin{array}{c}\nu^{\mu} \\ e^{\mu} \end{array}\right)_{L}$
	&	$0$	&$+$		\\
$q^{3}_{L}=\left(\begin{array}{c}u^{3} \\ d^{3} \end{array}\right)_{L}$
	&	$0$	&$+$	&&
$\ell^{\tau}_{L}=\left(\begin{array}{c}\nu^{\tau} \\ e^{\tau} \end{array}\right)_{L}$
	&	$-1$	&$+$	\\   \hline\hline

\begin{tabular}{c}$U_{R}^{1,3}$\\$U_{R}^{2}$\\$D_{R}^{1,2,3}$\end{tabular}	&	 
\begin{tabular}{c}$+2/3$\\$+2/3$\\$-1/3$\end{tabular}	&
\begin{tabular}{c}$+$\\$-$\\$-$\end{tabular}	&&
\begin{tabular}{c}$e_{R}^{e,\tau}$\\$e_{R}^{\mu}$\end{tabular}	&	
\begin{tabular}{c}$-4/3$\\$-1/3$\end{tabular}	&	
\begin{tabular}{c}$-$\\$-$\end{tabular}\\   \hline \hline 

\multicolumn{3}{|c}{Non-SM Quarks}	&&	\multicolumn{3}{c|}{Non-SM Leptons}	\\ \hline \hline
\begin{tabular}{c}$T_{L}$\\$T_{R}$\end{tabular}	&
\begin{tabular}{c}$+1/3$\\$+2/3$\end{tabular}	&
\begin{tabular}{c}$-$\\$-$\end{tabular}	&&
\begin{tabular}{c}$\nu_{R}^{e,\mu,\tau}$\\$\mathcal{N}_{R}^{e,\mu,\tau}$\end{tabular} 	&	
\begin{tabular}{c}$1/3$\\$0$\end{tabular}	&	
\begin{tabular}{c}$-$\\$-$\end{tabular}\\
$J^{1,2}_{L}$	&	  $0$ 	&$+$	&&	$E_{L},\mathcal{E}_{R}$	&	$-1$	&$+$	\\
$J^{1,2}_{R}$	&	 $-1/3$	&$+$	&&	$\mathcal{E}_{L},E_{R}$	&	$-2/3$	&$+$	\\ \hline \hline
\end{tabular}
\caption{Fermion particle content of the model, $X$ charge, $\mathbb{Z}_{2}$ parity and hypercharge. Here, $U^{1,2,3} =(u,c,t)$, $ D^{1,2,3}=(d,s,b)$, $e^{e,\mu,\tau}=(e,\mu,\tau)$ and $\nu^{e,\mu,\tau}=(\nu^{e},\nu^{\mu},\nu^{\tau})$. }
\label{fermionlist}
\end{table}

\subsection{Fermion masses}\label{fermions}

In the framework of Effective Field Theory (EFT), high energy physics effects are incorporated into the lower energy scale by integrating out the heavy degrees of freedom, which lead us to a dimensional expansion of an effective Lagrangian,
\begin{align}
    \mathcal{L}&=\mathcal{L}_{0} + \frac{\mathcal{L}_{1}}{\Lambda}+ \frac{\mathcal{L}_{2}}{\Lambda^{2}}+ ...,
\label{Leff}\end{align}
\noindent
where $\mathcal{L}_{0}$ contains all renormalizable interactions while $\mathcal{L}_{n}$, with $n\geq 1$, is a combination of non-renormalizable operators of dimension $n+4$ suppressed by powers of the NP energy scale $\Lambda^{n}$, in all cases restricted by gauge symmetry \cite{renorm}. The effective operators associated to the terms in the expansion encode loop processes that can be measured at a low energy scale, such as magnetic and electric dipole moments ($\bar{\psi}\sigma_{\mu\nu}\psi F^{\mu\nu}$ \cite{MDMEFT1} , $\bar{\psi}\sigma_{\mu\nu}\gamma^{5}\psi F^{\mu\nu}$ \cite{EDMEFT}), Higgs to di-photon decays ($hF_{\mu\nu}^{a}F^{a \; \mu\nu }$) \cite{htoggEFT1,*htoggEFT2,*htoggEFT3,*htoggEFT4,*htoggEFT5,*htoggEFT6} and particle masses \cite{EFTmass1,*EFTmass2,*EFTmass3,*EFTmass4,*EFTmass5,*EFTmass6}, which in general are sensitive to NP. Regarding the latter, certain mass matrix textures could at first leave massless the lightest fermions at tree-level, justifying then its smallness by loop processes. Equivalently, we can use the higher dimensional operators allowed by the symmetries of the model in the expansion Eq.(\ref{Leff}) and endow with masses the lightest fermions such as electron and up, down and strange quarks \cite{Froggatt:1978nt}.

\subsubsection{Charged lepton masses}\label{leptons}
The most general interaction Lagrangian involving charged leptons according to the $U(1)_{X} \otimes \mathbb{Z}_{2}$ symmetry is given by,
\small
\begin{align}
-\mathcal{L}_{\ell} &= 
\eta \overline{\ell^{e}_{L}}\phi_{2}e^{\mu}_{R} + h \overline{\ell^{\mu}_{L}}\phi_{2}e^{\mu}_{R} + 
\zeta\overline{\ell^{\tau}_{L}}\phi_{2}e^{e}_{R} + H\overline{\ell^{\tau}_{L}}\phi_{2}e^{\tau}_{R} +	
q_{11}\overline{\ell^{e}_{L}}\phi_{1}E_{R} + q_{21}\overline{\ell^{\mu}_{L}}\phi_{1}{E}_{R} +
g_{\chi E}\overline{E_{L}}\chi E_{R} + g_{\chi \mathcal{E}}\overline{\mathcal{E}_{L}}\chi^{*} \mathcal{E}_{R} + \mathrm{\text{H.C.}} . \label{lagE}
\end{align}
\normalsize
The charged $\mathcal{E}$ lepton is decoupled and gets a mass term $m_{\mathcal{E}}=g_{\chi\mathcal{E}}v_{\chi}/\sqrt{2}$, while after SSB, a $4\times 4$ mass matrix is obtained in the flavour basis $\mathbf{E}=(e^{e}, e^{\mu}, e^{\tau}, E)$ and can be written as,
\begin{equation}
\mathbb{M}_{E}^{0} = 
\left(\begin{array}{ c c c |c c}
    0 & \frac{\eta v_{2}}{\sqrt{2}} & 0 &  \frac{q_{11}v_{1}}{\sqrt{2}}   \\
    0 & \frac{h v_{2}}{\sqrt{2}}   & 0 &  \frac{q_{12}v_{1}}{\sqrt{2}}  \\
    \frac{\zeta v_{2}}{\sqrt{2}}  & 0 & \frac{H v_{2}}{\sqrt{2}} & 0 \\ \hline
    0 & 0 & 0 & \frac{g_{\chi E}v_{\chi}}{\sqrt{2}}
    \end{array} \right),
\end{equation}
and has rank 3, which means that the lightest lepton, the electron, is massless at tree-level. In view of this, as anticipated in the introduction, we consider the following effective operators up to dimension 7, invariant under the symmetry of the model, in order to endow the electron with mass,

\begin{align}
    \mathcal{O}^{\ell}_{ij}&=\Omega^{\ell}_{ij}\left(\frac{\chi^{*}}{\Lambda}\right)^{3}\bar{\ell}_{L}^{i}\phi_{2}e_{R}^{j},  & \mathcal{O}^{\ell}_{\tau\mu}&=\Omega^{\ell}_{32}\left(\frac{\chi}{\Lambda}\right)^{3}\bar{\ell}^{\tau}\phi_{2}e_{R}^{\mu}, \nonumber\\
    \mathcal{O}^{\ell}_{Ej}&= \frac{\Omega^{\ell}_{4j}}{\Lambda}(\phi_{2}^{\dagger}\phi_{1})\bar{E}_{L}e_{R}^{j}, & \mathcal{O}^{\ell}_{E\mu}&=\frac{\Omega^{\ell}_{42}}{\Lambda^{2}}(\phi_{1}^{\dagger}\phi_{2})\chi\bar{E}_{L}e_{R}^{\mu}, \nonumber \\
    \mathcal{O}^{\ell}_{\tau E}&=\Omega^{\ell}_{34}\left(\frac{\chi}{\Lambda}\right)^{3} \hat{\ell}_{L}^{\tau}\phi_{1} E_{R},
\end{align}
\noindent
where $i=e,\mu$, $j=e,\tau$ and $\Lambda$ is the associated energy scale. Thus, the new mass matrix reads,

\begin{equation}
\mathbb{M}_{E} = 
\left(\begin{array}{ c c c |c c}
    \Omega^{\ell}_{11}\frac{v_{2}v_{\chi}^{3}}{4\Lambda^{3}} & \frac{\eta v_{2}}{\sqrt{2}} & \Omega^{\ell}_{13}\frac{v_{2}v_{\chi}^{3}}{4\Lambda^{3}} &  \frac{q_{11}v_{1}}{\sqrt{2}}   \\
    \Omega^{\ell}_{21}\frac{v_{2}v_{\chi}^{3}}{4\Lambda^{3}}  & \frac{h v_{2}}{\sqrt{2}}   & \Omega^{\ell}_{23}\frac{v_{2}v_{\chi}^{3}}{4\Lambda^{3}} &  \frac{q_{12}v_{1}}{\sqrt{2}}  \\
    \frac{\zeta v_{2}}{\sqrt{2}}  & \Omega^{\ell}_{32} \frac{v_{2}v_{\chi}^{3}}{4\Lambda^{3}}                            & \frac{H v_{2}}{\sqrt{2}} &  \Omega^{\ell}_{34}\frac{v_{1}v_{\chi}^{3}}{4\Lambda^{3}} \\ \hline
    \frac{v_{1}v_{2}}{2\Lambda}\Omega^{\ell}_{41}& \frac{v_{1}v_{2}v_{\chi}}{2\sqrt{2}\Lambda^{2}}\Omega^{\ell}_{42} & \frac{v_{1}v_{2}}{2\Lambda}\Omega^{\ell}_{43} & \frac{g_{\chi E}v_{\chi}}{\sqrt{2}}
    \end{array} \right).\label{eq:ME_matrix}
\end{equation} \\

The diagonalization matrices from the flavour to the mass basis $\boldsymbol{e}=(e,\mu,\tau,E)$ for the left-handed leptons $\mathbb{V}^{E}_{L}$ and for the right-handed ones $\mathbb{V}^{E}_{R}$  are obtained by diagonalizing $\mathbb{M}_{E}\mathbb{M}_{E}^{\dagger}$ and $\mathbb{M}_{E}^{\dagger}\mathbb{M}_{E}$ respectively and are given by,
\begin{align}
\mathbf{E}_{L} &= 
\mathbb{V}^{E }_{L}
\mathbf{e}_{L}, &
\mathbf{E}_{R} &= 
\mathbb{V}^{E }_{R}
\mathbf{e}_{R},
\end{align}

\noindent
which, taking into account the hierarchy between the different VEVs, the texture of the mass matrix in Eq.(\ref{eq:ME_matrix}) and a universal see-saw mechanism, can be factorized approximately as

\begin{align}
\mathbb{V}^{E}_{L} &\approx\mathbb{V}^{E}_{L\; 1}\mathbb{V}^{E}_{L\; 2},& \mathbb{V}^{E}_{R} &\approx\mathbb{V}^{E}_{R\; 1}\mathbb{V}^{E}_{R\; 2},
\end{align}

where each matrix is defined in appendix \ref{chargedlpetons}. The corresponding mass eigenvalues will be given by,

 \begin{align}
     m_{e}^{2}&\approx \frac{v_{1}^{4} v_{2}^{2} }{4}  \Bigg(\frac{ s_{e\tau} v_{\chi}^{3} (\Omega^{\ell}_{23} s_{e\mu}-\Omega^{\ell}_{13} c_{e\mu})}{v_{1}^{2} \Lambda^3} +\frac{c_{e\tau} v_{\chi}^{3} (\Omega^{\ell}_{11} c_{e\mu}-\Omega^{\ell}_{21} s_{e\mu})}{ v_{1}^{2}\Lambda^3} + \frac{\left(q_{11} c_{e\mu}-q_{12} s_{e\mu}\right) \left(\Omega^{\ell}_{43} s_{e\tau}-\Omega^{\ell}_{41}c_{e\tau}\right)}{ \sqrt{2}m_{E} \Lambda} \Bigg)^{2}, \label{me}\\
     m_{\mu}^{2}&\approx\frac{1}{2}(\eta^{2} + h^{2})v_{2}^{2}, \\
     m_{\tau}^{2} &\approx \frac{1}{2} H^{2}v_{2}^{2},\label{eq:tau_mass} \\
     m_{E}^{2}&\approx \frac{1}{2}g_{\chi E}^{2}v_{\chi}^{2}.\label{CLmasses}
 \end{align}

\begin{figure}[ht]
    \centering
    \includegraphics[scale=0.25]{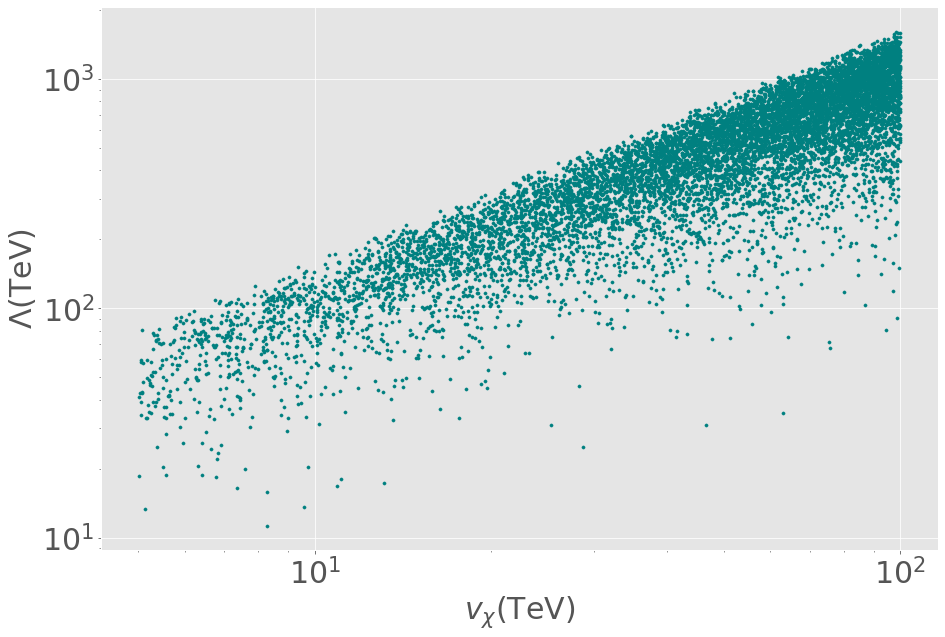}
    \caption{Monte Carlo exploration for $\Lambda$ as a function of $v_{\chi}$ according to the electron mass by exploring the parameter space of Eq.(\ref{me}).}
    \label{lambdavx}
\end{figure}

These mass eigenvalues impose some restrictions on the parameter space generated by $\{\eta, \zeta, g_{\chi E}, q_{11}, q_{12},\Omega_{ij}\}$. In particular, the mass of the top quark depends on $v_{1}$ while the bottom quark and the $\tau$ lepton masses depend on $v_{2}$. In this way, and consistently with the flavour fit of section \ref{Banom}, we can choose $v_{1}=245.6$ GeV and $v_{2}\approx 7$ GeV. On the other hand, the smallness of the electron mass is explained by the $\upsilon\Lambda^{-3}$ suppression in Eq.(\ref{me}), from which the limit $\Lambda \leq \sqrt[3]{9.4\frac{v_{2}}{4m_{e}}}v_{\chi}\approx 17 v_{\chi}$ (see figure \ref{lambdavx}) is obtained when scanning over the Yukawa couplings in the range $[1,\,10]$.

\subsubsection{Quark mixing}\label{quarks}
Since the new symmetry has a non-universal $X$ charge assignation, $Z_{2}^{\mu}$ flavour-changing neutral current interactions will be present in the Lagrangian. Then, prior to considering the relevant interaction Lagrangian for the $b \rightarrow s \ell^{+} \ell^{-}$ transition, we need to consider the rotation matrix that connects flavour and mass eigenstates in the quark sector. The most general Yukawa Lagrangian allowed by the $U(1)_X\otimes\mathbb{Z}_2$ symmetry is given by,

\begin{eqnarray}
-\mathcal{L}_U-\mathcal{L}_D &=& \overline{q_L^{1}}\left(\widetilde{\phi} _2h^{U}_2 \right)_{12}U_R^{2}+\overline{q_L^{2}}(\widetilde{\phi }_1 h^{U}_{1})_{22}U_R^{2} +\overline{q_L^{3}}(\widetilde{\phi }_1 h^{U}_{1})_{31}U_R^{1}+\overline{q_L^{3}}(\widetilde{\phi }_1 h^{U}_{1})_{33}U_R^{3} \nonumber \\
&+& \overline{T_{L}}\left(\chi h_{\chi }^{U}\right)_{2}{U}_{R}^{2}+\overline{T_{L}}\left( \chi h_{\chi }^{T}\right){T}_{R} + \overline{q_L^{1}}\left(\widetilde{\phi} _2 h^{T}_{2} \right)_1T_R +\overline{q_L^{2}} (\widetilde{\phi } _1 h^{T}_{1})_2 T_R, \\
 &+& \overline{q_L^{3}}\left(\phi _2 h^{D}_{2} \right)_{3j}D_R^{j} \notag + \overline{q_L^{1}} (\phi _1 h^{J}_{1})_{m} J^{m}_R + \overline{q_L^{2}}\left(\phi  _2 h^{J}_{2} \right)_{m} J^{m}_R+\overline{J_{L}^n}\left(\chi ^*h_{\chi }^{J}\right)_{nm}{J}_{R}^{m}+\text{h.c.},
\end{eqnarray}

\noindent
where $\tilde{\phi}_{1,2}=i\sigma_{2}\phi_{1,2}^{*}$ are conjugate fields, $j = 1,2,3$ label the right-handed fermions and $n(m) = 1, 2$ is the index of the exotic $\mathcal{J}^{n(m)}$ quarks. After SSB takes place, such Lagrangians give rise to the following mass matrices,

\small
\begin{align}
\mathbb{M}_{U}^{0} & =\left(\begin{array}{ccc|cc}
0 & \frac{(h_{2}^{U})_{12}v_{2}}{\sqrt{2}} & 0 & \frac{(h_{2}^{T})_{1}v_{2}}{\sqrt{2}}\\
0 & \frac{(h_{1}^{U})_{22}v_{1}}{\sqrt{2}} & 0 & \frac{(h_{1}^{T})_{2}v_{1}}{\sqrt{2}}\\
\frac{(h_{1}^{U})_{13}v_{1}}{\sqrt{2}} & 0 & \frac{(h_{1}^{U})_{33}v_{1}}{\sqrt{2}} & 0\\
\hline 0 & \frac{(h_{\chi}^{U})_{2}v_{\chi}}{\sqrt{2}} & 0 & \frac{h_{\chi}^{T}v_{\chi}}{\sqrt{2}}
\end{array}\right), & \mathbb{M}_{D}^{0} & =\left(\begin{array}{ccc|cc}
0 & 0 & 0 & \frac{v_{1}(h_{1}^{J})_{1}}{\sqrt{2}} & \frac{v_{1}(h_{1}^{J})_{2}}{\sqrt{2}}\\
0 & 0 & 0 & \frac{v_{2}(h_{2}^{J})_{1}}{\sqrt{2}} & \frac{v_{2}(h_{2}^{J})_{2}}{\sqrt{2}}\\
\frac{v_{2}(h_{2}^{D})_{31}}{\sqrt{2}} & \frac{v_{2}(h_{2}^{D})_{32}}{\sqrt{2}} & \frac{v_{2}(h_{2}^{D})_{33}}{\sqrt{2}} & 0 & 0\\
\hline 0 & 0 & 0 & \frac{v_{\chi}(h_{\chi}^{J})_{11}}{\sqrt{2}} & \frac{v_{\chi}(h_{\chi}^{J})_{12}}{\sqrt{2}}\\
0 & 0 & 0 & \frac{v_{\chi}(h_{\chi}^{J})_{21}}{\sqrt{2}} & \frac{v_{\chi}(h_{\chi}^{J})_{22}}{\sqrt{2}}
\end{array}\right).\label{quarkmatrices}
\end{align}
\normalsize
In the case of up-like quarks, the mass matrix has rank 3 which means that the up quark is massless.  Similarly to the electron case, we can consider the following set of dimension 5 effective operators,

\begin{align*}
    \mathcal{O}_{11}^{U}&=\Omega^{U}_{11}\frac{\chi^{*}}{\Lambda}\bar{q}_{L}^{1}\tilde{\phi}_{1}U_{R}^{1}, & \mathcal{O}_{13}^{U}&=\Omega^{U}_{13}\frac{\chi^{*}}{\Lambda}\bar{q}_{L}^{1}\tilde{\phi}_{1}U_{R}^{3}, \\
    \mathcal{O}_{21}^{U}&=\Omega^{U}_{21}\frac{\chi}{\Lambda}\bar{q}_{L}^{2}\tilde{\phi}_{2}U_{R}^{1}, &  \mathcal{O}_{23}^{U}&=\Omega^{U}_{23}\frac{\chi}{\Lambda}\bar{q}_{L}^{2}\tilde{\phi}_{2}U_{R}^{1}, \\
    \mathcal{O}_{32}^{U}&=\Omega^{U}_{32}\frac{\chi}{\Lambda}\bar{q}_{L}^{3}\tilde{\phi}_{2}U_{R}^{2}, &  \mathcal{O}_{34}^{U}&=\Omega^{U}_{34}\frac{\chi}{\Lambda}\bar{q}_{L}^{3}\tilde{\phi}_{2}\mathcal{T}_{R}, \\
    \mathcal{O}_{41}^{U}&=\Omega^{U}_{41}\frac{\phi_{1}^{\dagger}\phi_{2}}{\Lambda}\bar{\mathcal{T}}_{L}U_{R}^{1}, & \mathcal{O}_{43}^{U}&=\Omega^{U}_{43}\frac{\phi_{1}^{\dagger}\phi_{2}}{\Lambda}\bar{\mathcal{T}}_{L}U_{R}^{3},
\end{align*}

in such a way that all zeros in the mass matrix are filled, yielding the following mass matrix,

\begin{equation}
\mathbb{M}_{U} = 
\left(\begin{array}{ c c c |c c}
    \Omega^{U}_{11}v_{1}\frac{v_{\chi}}{2\Lambda} & \frac{(h_{2}^{U})_{12} v_{2}}{\sqrt{2}} & \Omega^{U}_{13}v_{1}\frac{v_{\chi}}{2\Lambda} &  \frac{(h_{2}^{T})_{1}v_{2}}{\sqrt{2}}   \\
    \Omega^{U}_{21}v_{2}\frac{v_{\chi}}{2\Lambda} & \frac{(h_{1}^{U})_{22} v_{1}}{\sqrt{2}}   & \Omega^{U}_{23}v_{2}\frac{v_{\chi}}{2\Lambda} &  \frac{(h_{1}^{T})_{2}v_{1}}{\sqrt{2}}  \\
    \frac{(h_{1}^{U})_{13} v_{1}}{\sqrt{2}}  & \Omega^{U}_{32}v_{2}\frac{v_{\chi}}{2\Lambda} & \frac{(h_{1}^{U})_{33} v_{1}}{\sqrt{2}} & \Omega^{U}_{34}v_{2}\frac{v_{\chi}}{2\Lambda} \\ \hline
    \Omega^{U}_{41}\frac{v_{1}v_{2}}{2\Lambda} & \frac{(h_{\chi}^{U})_{2} v_{\chi}}{\sqrt{2}}& \Omega^{U}_{43}\frac{v_{1}v_{2}}{2\Lambda} & \frac{h_{\chi}^{T}v_{\chi}}{\sqrt{2}}
    \end{array} \right).
\end{equation}

In order to provide mass eigenvalues and rotation matrices for the up-like quarks, as in the charged leptons case, the left-handed quark rotation matrix can be written as  $\mathbb{V}^{U}_{L}\approx\mathbb{V}^{U}_{L\; 1}\mathbb{V}^{U}_{L\; 2}$ while for right-handed quarks we have a single matrix $\mathbb{V}^{U}_{R}$. The explicit expressions for the matrices are given by in appendix \ref{upquarks}. The mass eigenvalues are given by,
 \begin{align}
m_{u}^{2}\approx &\frac{v_{\chi}^{2}}{4\Lambda^{2}}\Big(v_{1}s_{uc}(\Omega_{13}^{U}s_{ut}-\Omega_{11}^{U}c_{ut}) + v_{2}c_{uc}(-\Omega_{23}^{U}s_{ut}+\Omega_{21}^{U}c_{ut})\Big)^{2}, \label{mup}\\
m_{c}^{2}\approx&\frac{1}{2}(v_{2}^{2}r_{1}^{+2} + v_{1}^{2}r_{2}^{+2}), \label{mc} \\
m_{t}^{2}\approx&\frac{1}{2}v_1^2\left[((h_{1}^{U})_{13})^2+((h_{1}^{U})_{33})^2\right],\\
m_{T}^{2}\approx&\frac{1}{2}v_\chi^2\left[(h_{\chi}^{T})^2+((h_{\chi}^{U})_{2})^2\right].
 \end{align}

The mass matrix for the up-like quarks is analogous to the one for the charged leptons. We obtained that both muon and tau leptons are proportional to $v_{2}$, where the quotient $m_{\mu}/m_{\tau} = 0.059$ can be understood with Yukawa couplings of order 1. Although, in the case of up-like quarks, having $m_{c}, m_{t} \propto v_{1} $ is not possible due to the large mass difference between the top and charm quarks. The mass matrix entry $(\mathbb{M}_{U})_{42}=(h_{\chi}^{U})_{2}v_{\chi}/\sqrt{2}$ together with the $(\mathbb{M}_{U})_{24}$ one, contributes to the SM quark mixing through a see-saw mechanism between $c$ and $\mathcal{T}$ quarks, producing a Yukawa difference, as shown in the charm mass expression in Eq. (\ref{mc}) that can be approximated by,
\begin{align}
    m_{c}^{2} & \approx \frac{1}{2}v_{1}^{2}r_{2}^{+ 2} \\
    &= \frac{1}{2}v_{1}^{2} \frac{( (h_{1}^{T})_{2}(h_{\chi}^{U})_{2} - (h_{1}^{U})_{22}h_{\chi}^{T})^{2}}{ (h_{\chi}^{U})_{2}^{2} + (h_{\chi}^{T})^{2} }.
\end{align}
Since all Yukawa couplings are of order 1, it would give a mass to the charm quark proportional to the top quark, however a suppression factor $(h_{1}^{T})_{2}c_{\alpha} - (h_{1}^{U})_{22}s_{\alpha}$ sets the order of magnitude to $10^{-2}$, consistent with the measured mass values,

\begin{align}
    \frac{m_{c}}{m_{t}} \approx \frac{ (h_{1}^{T})_{2}c_{\alpha} - (h_{1}^{U})_{22}s_{\alpha}}{ \sqrt{((h_{1}^{U})_{13})^{2} + ((h_{1}^{U})_{33})^{2} } } \sim 10^{-2}.
\end{align}

\begin{figure}[ht]
    \centering
    \includegraphics[scale=0.25]{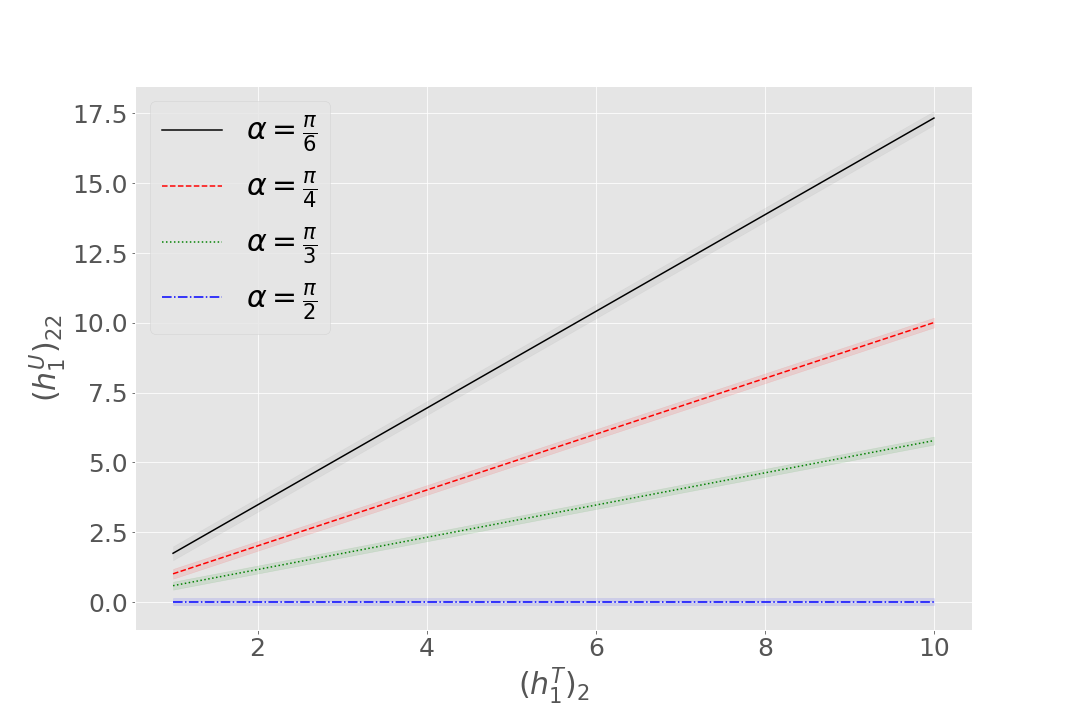}
    \caption{Parameter region compatible with the charm and top masses.}
    \label{ctmasses}
\end{figure}

As a benchmark scenario, taking $(h_{1}^{U})_{13} = (h_{1}^{U})_{33} = 1/\sqrt{2}$, and with $m_{c}=1.280 \pm 0.025$ GeV\cite{CKMparams} and $m_{t}=172.69 \pm 0.48$ GeV\cite{topmass}, the ratio of these quarks masses becomes,

\begin{align}
    (7.33 \pm 0.124)\times10^{-3}\approx  (h_{1}^{T})_{2}c_{\alpha} - (h_{1}^{U})_{22}s_{\alpha}.
\end{align}
\noindent
This requirement can be easily achieved as it can be seen in figure \ref{ctmasses}, where we show the parameter space compatible with that constraint for different values of $\alpha$.

\begin{figure}[ht]
    \centering
    \includegraphics[scale=0.25]{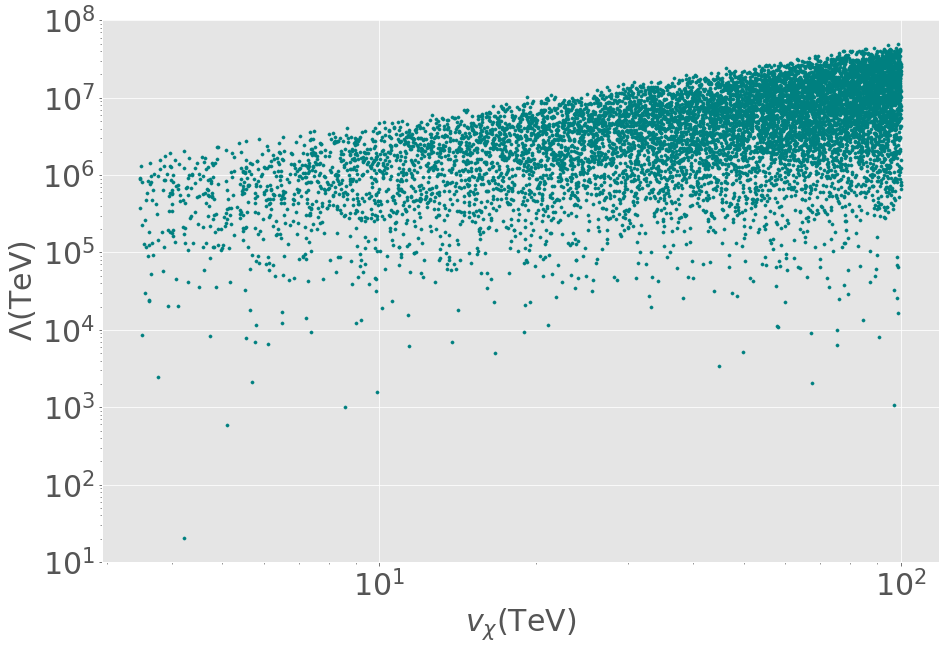}
    \caption{Monte Carlo exploration for $\Lambda$ as a function of $v_{\chi}$  according to the up quark mass in Eq.(\ref{mup}).}
    \label{LambdaU}
\end{figure}

We can also get an estimate of the higher dimensional operators energy scale from the up quark mass. The Monte Carlo scan gives uniformly distributed random values to the parameter space in Eq.(\ref{mup}) prior to numerically solving the equation for $\Lambda$. It shows that there is also an upper limit of $\Lambda \leq \frac{5 v_{1}}{m_{u}}v_{\chi}\approx 5.6\times 10^{5} v_{\chi}$ as it can be seen in figure \ref{LambdaU}. Such an upper limit is several orders of magnitude larger than the one obtained from the electron mass despite having similar mass values. This can be understood by noticing that the effective operators that contribute to the electron mass are of dimension seven while effective operators responsible of the up quark mass are of dimension five, requiring then larger values of $\Lambda$.

Likewise, the down-like quark mass matrix written in the basis $\{d_{1},d_{2},d_{3},\mathcal{J}^{2},\mathcal{J}^{2}\}$, as shown in Eq.(\ref{quarkmatrices}), has rank 3. In this case, the two lightest particles, namely the down and strange quarks, are massless. To alleviate this, we consider the following set of dimension 5 effective operators,

\begin{align}
    \mathcal{O}_{1j}^{D}&=\Omega_{1j}^{D}\frac{\chi^{*}}{\Lambda}\bar{q}_{L}^{1}\phi_{2}D_{R}^{j}, & \mathcal{O}_{2j}^{D}&=\Omega_{2j}^{D}\frac{\chi}{\Lambda}\bar{q}_{L}^{2}\phi_{1}D_{R}^{j}, \\
    \mathcal{O}_{4j}^{D}&=\Omega_{4j}^{D}\frac{\phi_{2}^{\dagger}\phi_{1}}{\Lambda}\bar{\mathcal{J}}_{L}^{1}D_{R}^{j}, & \mathcal{O}_{5j}^{D}&=\Omega_{5j}^{D}\frac{\phi_{2}^{\dagger}\phi_{1}}{\Lambda}\bar{\mathcal{J}}_{L}^{2}D_{R}^{j}, \\
    \mathcal{O}_{34}^{D}&=\Omega_{34}^{D}\frac{\chi}{\Lambda}\bar{q}_{L}^{3}\phi_{2}\mathcal{J}_{R}^{1}, & \mathcal{O}_{35}^{D}&=\Omega_{35}^{D}\frac{\chi}{\Lambda}\bar{q}_{L}^{3}\phi_{2}\mathcal{J}_{R}^{2},
\end{align}
filling out all zeros as in the quarks and charged leptons cases. The resulting mass matrix is,
 
\begin{equation}
\mathbb{M}_{D}=
\left(\begin{array}{ccc|cc}
\Omega_{11}^{D}v_{2}\frac{v_{\chi}}{2\Lambda} & \Omega_{12}^{D}v_{2}\frac{v_{\chi}}{2\Lambda} & \Omega_{13}^{D}v_{2}\frac{v_{\chi}}{2\Lambda} & \frac{v_{1}(h_{1}^{J})_{1}}{\sqrt{2}} & \frac{v_{1}(h_{1}^{J})_{2}}{\sqrt{2}}\\
\Omega_{21}^{D}v_{1}\frac{v_{\chi}}{2\Lambda} & \Omega_{22}^{D}v_{1}\frac{v_{\chi}}{2\Lambda} & \Omega_{23}^{D}v_{1}\frac{v_{\chi}}{2\Lambda} & \frac{v_{2}(h_{2}^{J})_{1}}{\sqrt{2}} & \frac{v_{2}(h_{2}^{J})_{2}}{\sqrt{2}}\\
\frac{v_{2}(h_{2}^{D})_{31}}{\sqrt{2}} & \frac{v_{2}(h_{2}^{D})_{32}}{\sqrt{2}} & \frac{v_{2}(h_{2}^{D})_{33}}{\sqrt{2}} & \Omega_{34}^{D}v_{2}\frac{v_{\chi}}{2\Lambda} & \Omega_{35}^{D}v_{2}\frac{v_{\chi}}{2\Lambda}\\
\hline \Omega_{41}^{D}v_{2}\frac{v_{\chi}}{2\Lambda} & \Omega_{42}^{D}v_{2}\frac{v_{\chi}}{2\Lambda} & \Omega_{43}^{D}v_{2}\frac{v_{\chi}}{2\Lambda} & \frac{v_{\chi}(h_{\chi}^{J})_{11}}{\sqrt{2}} & \frac{v_{\chi}(h_{\chi}^{J})_{12}}{\sqrt{2}}\\
\Omega_{51}^{D}v_{2}\frac{v_{\chi}}{2\Lambda} & \Omega_{52}^{D}v_{2}\frac{v_{\chi}}{2\Lambda} & \Omega_{53}^{D}v_{2}\frac{v_{\chi}}{2\Lambda} & \frac{v_{\chi}(h_{\chi}^{J})_{21}}{\sqrt{2}} & \frac{v_{\chi}(h_{\chi}^{J})_{22}}{\sqrt{2}}
\end{array}\right),
\end{equation}
\noindent
where mass eigenvalues can be written as,
\small
\begin{align}
m_{d}^{2}&=\frac{v_{\chi}^2\left(\xi_{22} v_1^2+\xi_{11} v_2^2-\sqrt{4 \xi_{12}^2 v_1^2 v_2^2+\left(\xi_{22} v_1^2-\xi_{11} v_2^2\right){}^2}\right) }{8 \Lambda^2}, \label{mdown}\\
m_{s}^{2}&=\frac{v_{\chi}^2\left(\xi_{22} v_1^2+\xi_{11} v_2^2+\sqrt{4 \xi_{12}^2 v_1^2 v_2^2+\left(\xi_{22} v_1^2-\xi_{11} v_2^2\right){}^2}\right) }{8 \Lambda^2},\label{mstrange} \\
m_{b}^{2}&=\frac{1}{2}v_{2}^{2}(((h_{2}^{D})_{31})^{2}+((h_{2}^{D})_{32})^{2}+((h_{2}^{D})_{33})^{2}), \\
m_{\mathcal{J}^{1}}&=\frac{1}{4} v_{\chi}^2 \left(\rho-\sqrt{\rho^2-4 \eta^2}\right), \\
m_{\mathcal{J}^{2}}&=\frac{1}{4} v_{\chi}^2\left(\rho+\sqrt{\rho^2-4 \eta^2}\right),\label{mJ2}
\end{align}
\normalsize

with the parameters $xi_{ij}$, $\rho$ and $\eta$ defined in appendix \ref{downquarks}. \\

\begin{figure}
    \centering
    \includegraphics[scale=0.25]{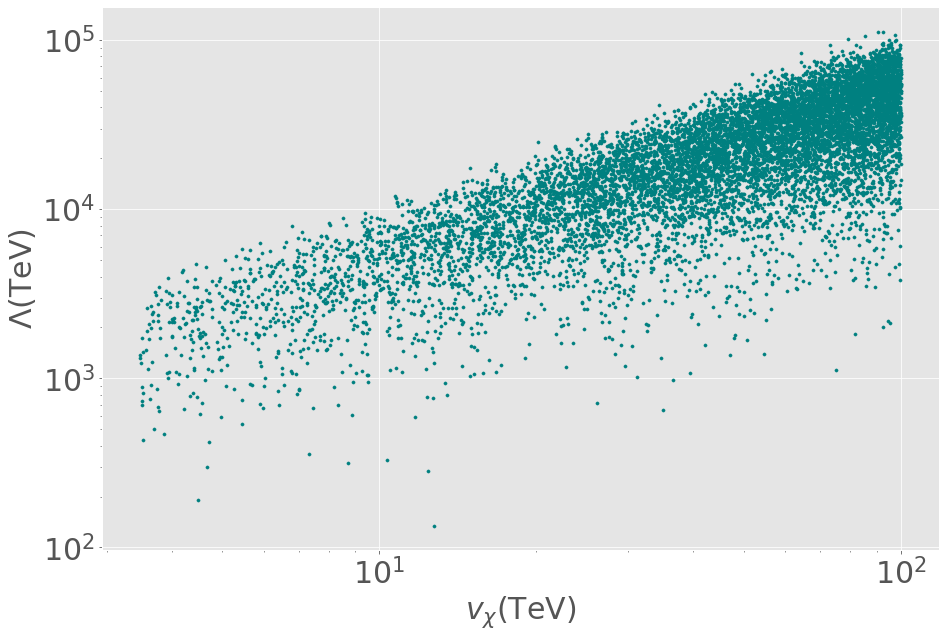}
    \caption{Monte Carlo exploration for $\Lambda$ as a function of $v_{\chi}$ according to the down and strange masses in Eq.(\ref{mdown}) and Eq.(\ref{mstrange}) respectively.}
    \label{LambdaD}
\end{figure}

Given that we have two particle masses generated by effective operators, we assign random values evenly distributed to all $\Omega^{D}_{ij}$ parameters  in the interval $[1,10]$ excepting for $\Omega_{11}^{D}$ given that along $\Lambda$, are unknown parameters fixing both the down and strange quark masses. The allowed values for $\Lambda$ as a function of $v_{\chi}$ are shown in figure \ref{LambdaD}, where an upper bound of $\Lambda \leq \left(\frac{v_{2}}{2\sqrt{2}m_{d}}0.063\right)v_{\chi}\approx 4.7v_{\chi}$ is obtained.

\begin{table}[ht]
    \centering
    \begin{tabular}{|c|c|} \hline
        Fermion &  $\Lambda$ upper bound \\ \hline
        $e$ &  $\Lambda \leq \sqrt[3]{9.4\frac{v_{2}}{4m_{e}}}\approx 17 v_{\chi}$ \\ \hline
        $u$ & $\Lambda \leq \frac{5 v_{1}}{m_{u}} \approx 5.6\times 10^{5}v_{\chi}$ \\ \hline
        $d,s$ & $\Lambda \leq \left(\frac{v_{2}}{2\sqrt{2}m_{d}}0.063\right)\approx  4.7 v_{\chi}$ \\ \hline
     \end{tabular}
    \caption{$\Lambda$ scale upper bound according to each light fermion mass.}
    \label{Lambda-bounds}
\end{table}

The upper bounds for $\Lambda$ are summarized in table \ref{Lambda-bounds}. It can be seen that the down quark sector yields the smallest upper bound,  meaning that SM fermion masses restrict the effective operators energy scale to $\Lambda \leq 4.7 v_{\chi}$.

\section{Muon \texorpdfstring{$g-2$}{} anomaly}\label{g-2anom}

In view that the model considers the existence of several new particles such as charged scalars and heavy Majorana neutrinos, their contributions to muon $g-2$ can be considered as well. The different one-loop diagrams that might contribute are shown in figure \ref{g-2feynman}.

\begin{figure}[ht]
    \LARGE
     \centering
     \begin{subfigure}[b]{0.2\textwidth}
         \centering
             \begin{tikzpicture}
  \begin{feynman}
    \vertex (a) ;
    \vertex [below right = of a] (b);
    \vertex [below left =of a] (c) ;
    \vertex [right = of a] (x){\( f\)};
    \vertex [left =of a] (y){\( f\)} ;
    \vertex [below = of a ](z){\( Z_{2} \)};
    \vertex [above = of a ] (d){\( \gamma \) };
    \vertex [below right = of b] (e){\(\mu \) } ;
    \vertex [below left = of c] (f){\( \mu \) } ;
    \diagram* {
      (a) -- [anti fermion ] (b) -- [boson] (c) -- [anti fermion] (a),
      (a) -- [boson] (d),
      (b) -- [ anti fermion] (e),
      (c) -- [ fermion] (f),
    };
  \end{feynman}
\end{tikzpicture}
         \caption{}
         \label{g-2a}
     \end{subfigure}
     \hfill
      \begin{subfigure}[b]{0.2\textwidth}
    \centering
    \begin{tikzpicture}
  \begin{feynman}
    \vertex (a) ;
    \vertex [below right = of a] (b);
    \vertex [below left =of a] (c) ;
    \vertex [right = of a] (x){\( f\)};
    \vertex [left =of a] (y){\( f \)} ;
    \vertex [below = of a ](z){\( H,A_0 \)};
    \vertex [above = of a ] (d){\( \gamma \) };
    \vertex [below right = of b] (e){\( \mu \) } ;
    \vertex [below left = of c] (f){\( \mu \) } ;
    \diagram* {
      (a) -- [anti fermion ] (b) -- [scalar] (c) -- [anti fermion] (a),
      (a) -- [boson] (d),
      (b) -- [anti fermion] (e),
      (c) -- [ fermion] (f),
    };
  \end{feynman}
\end{tikzpicture}
         \caption{}
         \label{g-2b}
     \end{subfigure}
     \hfill
          \begin{subfigure}[b]{0.2\textwidth}
         \centering
             \begin{tikzpicture}
  \begin{feynman}
    \vertex (a) ;
    \vertex [below right = of a] (b);
    \vertex [below left =of a] (c) ;
    \vertex [right = of a] (x){\( W^+ \)};
    \vertex [left =of a] (y){\( W^- \)} ;
    \vertex [below = of a ](z){\( N \)};
    \vertex [above = of a ] (d){\( \gamma \) };
    \vertex [below right = of b] (e){\( \mu \) } ;
    \vertex [below left = of c] (f){\( \mu \) } ;
    \diagram* {
      (a) -- [boson ] (b) -- [fermion] (c) -- [boson] (a),
      (a) -- [boson] (d),
      (b) -- [ anti fermion] (e),
      (c) -- [fermion] (f),
    };
  \end{feynman}
\end{tikzpicture}
         \caption{}
         \label{g-2c}
     \end{subfigure}
     \hfill
     \begin{subfigure}[b]{0.2\textwidth}
         \centering
             \begin{tikzpicture}
  \begin{feynman}
    \vertex (a) ;
    \vertex [below right = of a] (b);
    \vertex [below left =of a] (c) ;
    \vertex [right = of a] (x){\( H^+ \)};
    \vertex [left =of a] (y){\( H^- \)} ;
    \vertex [below = of a ](z){\( N \)};
    \vertex [above = of a ] (d){\( \gamma \) };
    \vertex [below right = of b] (e){\( \mu \) } ;
    \vertex [below left = of c] (f){\( \mu \) } ;
    \diagram* {
      (a) -- [scalar ] (b) -- [fermion] (c) -- [scalar] (a),
      (a) -- [boson] (d),
      (b) -- [ anti fermion] (e),
      (c) -- [fermion] (f),
    };
  \end{feynman}
\end{tikzpicture}
         \caption{}
         \label{g-2d}
     \end{subfigure}
    \caption{Contribution to muon $g-2$ from the interaction to the $Z_2$ neutral gauge boson (a),  (pseudo)scalars (b), charged $W^{+}$ gauge boson with exotic neutrinos (c) and charged scalars with neutrinos (d).}
    \label{g-2feynman}
\end{figure}

The most general interaction Lagrangian involving neutral leptons is,

\begin{align}
    \mathcal{L}_{NL}&= h_{2p}^{\nu w} \bar{\ell}_{L}^{p}\tilde{\phi}_ {2}\nu_{R}^{w} + h_{\chi q}^{\nu j}\bar{\nu}_{R}^{w\; c}\chi^{*}N_{R}^{j} + \frac{1}{2}\bar{N}_{R}^{i\; c}M_{N}^{ij}N_{R}^{j}\label{lagN},
\end{align}
\noindent
where $p=e,\mu$ labels the lepton doublets, $w=e,\mu,\tau$ labels right-handed neutrinos and $i,j=e,\mu,\tau$ labels the Majorana neutrinos (see appendix \ref{neutrinomasses}). Such Lagrangian is responsible of neutrino mass generation via an inverse see-saw mechanism and is able to reproduce the PMNS matrix as shown in \cite{orig,modelPMNS}. In order to explore the general behavior of the model, we consider a benchmark scenario able to reproduce neutrino masses and PMNS matrix, identified by,

\begin{align}
     r_{1}&= 3.5\times10^{-3}, &  r_{2}&= 1.08\times10^{-3}, \nonumber\\
    h_{2e}^{\nu e}&= 4.08 e^{-0.129i}, &    h_{2\mu}^{\nu e}&=-2.28,   \nonumber \\
   h_{2e}^{\nu \mu}&= 3.38e^{0.216i},   &  h_{2\mu}^{\nu \mu}&= 0.48,  \nonumber\\
  h_{2e}^{\nu \tau}&=4.70e^{0.0103i}, & h_{2\mu}^{\nu \tau}&=1.80, \nonumber\\
    \theta_{e\mu}&=0.997. \label{params}
\end{align}
Besides this, we consider the case where exotic neutrinos have nearly degenerate masses, $m_{\mathcal{N}_{i}} \approx m_{\mathcal{N}_{j}}$ for $i,j=1,...,6$, so their masses are given by the single mass parameters $m_{\mathcal{N}}$ (see Appendix \ref{neutrinomasses}). Additionally, the lepton couplings $q_{11}$ and $q_{22}$ introduced in Eq. (\ref{lagE}) have a negligible effect on the lepton phenomenology because their effect on the charged lepton masses only is suppressed as can be seen in Eq. (\ref{CLmasses}). Furthermore, their effect on the PMNS matrix is given by the charged lepton rotation, where $q_{11}$ and $q_{12}$ are involved in the see-saw decoupling of the exotic $E$ lepton and therefore suppressed by $m_{E}$. In this way, at first $q_{11}$ and $q_{12}$ could take any values between 1 and 10, specifically we choose  $q_{11}=5.957$ and $q_{12}=7.373$.

\subsection{\texorpdfstring{$Z_2$}{} contributions}
Such an interaction produces a contribution to muon $g-2$ at one-loop level according to figure \ref{g-2a}. The interaction terms between muon, $E$ and $Z_{2}$ can can be written as,

\begin{align}
    \mathcal{L}&=ig \bar{E}\slashed{Z}_{2} [(  (J_{\tau}^{L2} - J_{s'}^{L2})(\mathbb{V}_{L}^{E \dagger})^{E\tau}(\mathbb{V}_{L}^{E})^{\tau \mu} + (J_{E}^{L2}-J_{s'}^{L2})(\mathbb{V}_{L}^{E \dagger})^{EE}(\mathbb{V}_{L}^{E})^{E \mu}) )P_{L}\nonumber \\
    &+ ((J_{\mu}^{R2} - J_{p}^{R2})(\mathbb{V}_{R}^{E\dagger})^{E\mu }(\mathbb{V}_{R}^{E})^{\mu \mu} + (J_{E}^{R2}-J_{p}^{R2})(\mathbb{V}_{R}^{E \dagger})^{E E}(\mathbb{V}_{R}^{E})^{E \mu}) )P_{R}]\mu.
\end{align}    
    
The rotation matrices indicate that $(\mathbb{V}_{R}^{E})^{\mu \mu}=1$, but $(\mathbb{V}_{R}^{E})^{E\mu } \propto m_{E}^{-2}$, so we can neglect all the right-handed couplings. Furthermore, in the left-handed lepton rotation matrices we have $(\mathbb{V}_{L}^{E})^{EE}=1$ and $(\mathbb{V}_{L}^{E})^{E \tau } = 0$, so we keep only the second line which is of the order $\mathcal{O}(m_{E}^{-1})$, resulting in \cite{g-2},

\small
\begin{align}
    \mathcal{L}&=ig \bar{E}\slashed{Z}_{2} [ (J_{E}^{L2}-J_{s'}^{L2})(\mathbb{V}_{L}^{E \dagger})^{EE}(\mathbb{V}_{L}^{E})^{E \mu}) )P_{L}]\mu +\text{h.c.}\nonumber \\
    &=ig \bar{E}\slashed{Z}_{2} [ \left(\frac{s_{Z}}{2c_{W}} + \frac{g_{X}}{g}c_{Z}\right)(\mathbb{V}_{L}^{E \dagger})^{EE}(\mathbb{V}_{L}^{E})^{E \mu}) )P_{L}]\mu +\text{h.c.}\nonumber \\
    & \approx -i\frac{v_{1}(s_{e\mu}q_{11}+c_{e\mu}q_{12})}{\sqrt{2}m_{E}}\left(\frac{g}{2c_{W}}s_{Z} + g_{X}\right)\bar{E}\slashed{Z}_{2}P_{L}\mu  + \text{h.c.}.\label{Z2leptons}
\end{align}
\normalsize

The general expression for muon $g-2$ due to a mediating neutral gauge boson can be found in \cite{g-2},

\begin{align}
\Delta a_{\mu}^{Z_{2}} &= \frac{1}{8\pi^2}\frac{m_\mu^2}{ M_{Z_{2}}^2 } \int_0^1 dx \frac{g_{v}^2 \ P_{v}(x) + g_{a}^2 \ P_{a} (x) }{(1-x)(1-\lambda^{2} x) +\epsilon^2 \lambda^2 x},
\end{align}
where 
\small
\begin{align}
P_{v}(x) & =   2x(1-x)(x-2(1-\epsilon))+\lambda^2(1-\epsilon)^2x^2(1+\epsilon-x), \nonumber\\
P_{a}(x) & =   2x(1-x)(x-2(1+\epsilon))+\lambda^2(1+\epsilon)^2x^{2}(1-\epsilon -x),
\end{align}
\normalsize
\noindent
with $\epsilon = m_{E}/m_{\mu}$, $\lambda= m_{\mu}/M_{Z_{2}}$, and $g_{v}$ and $g_{a}$ are the vector and axial couplings respectively from the $\bar{E}\slashed{Z}_{2}P_{L}\mu$ vertex in Eq.(\ref{Z2leptons}), which obeys $g_{v}=-g_{a}\equiv g_{Z_{2}}$ defined as,
\begin{align}
   g_{Z_{2}} &\approx -\frac{v_{1}g_{X}(s_{e\mu}q_{11}+c_{e\mu}q_{12})}{\sqrt{2}m_{E}},
\end{align}

and $\Delta a_{\mu}^{Z_{2}}$ can be written as,

\begin{align}
\Delta & a_{\mu}^{Z_{2}} = \frac{9}{4\pi^2}\frac{m_\mu^2}{ v_{\chi}^2 } \left( \frac{v_{1}(s_{e\mu}q_{11}+c_{e\mu}q_{12})}{\sqrt{2}m_{E}}\right)^{2} \int_0^1 dx \frac{ 2x(1-x)(x-2) +\lambda^2 x^2 (1 - x - \epsilon^2 (1 + x)) }{(1-x)(1-\lambda^{2} x) +\epsilon^2 \lambda^2 x}.
\end{align}

\begin{figure}[ht]
    \centering
    \includegraphics[scale=0.25]{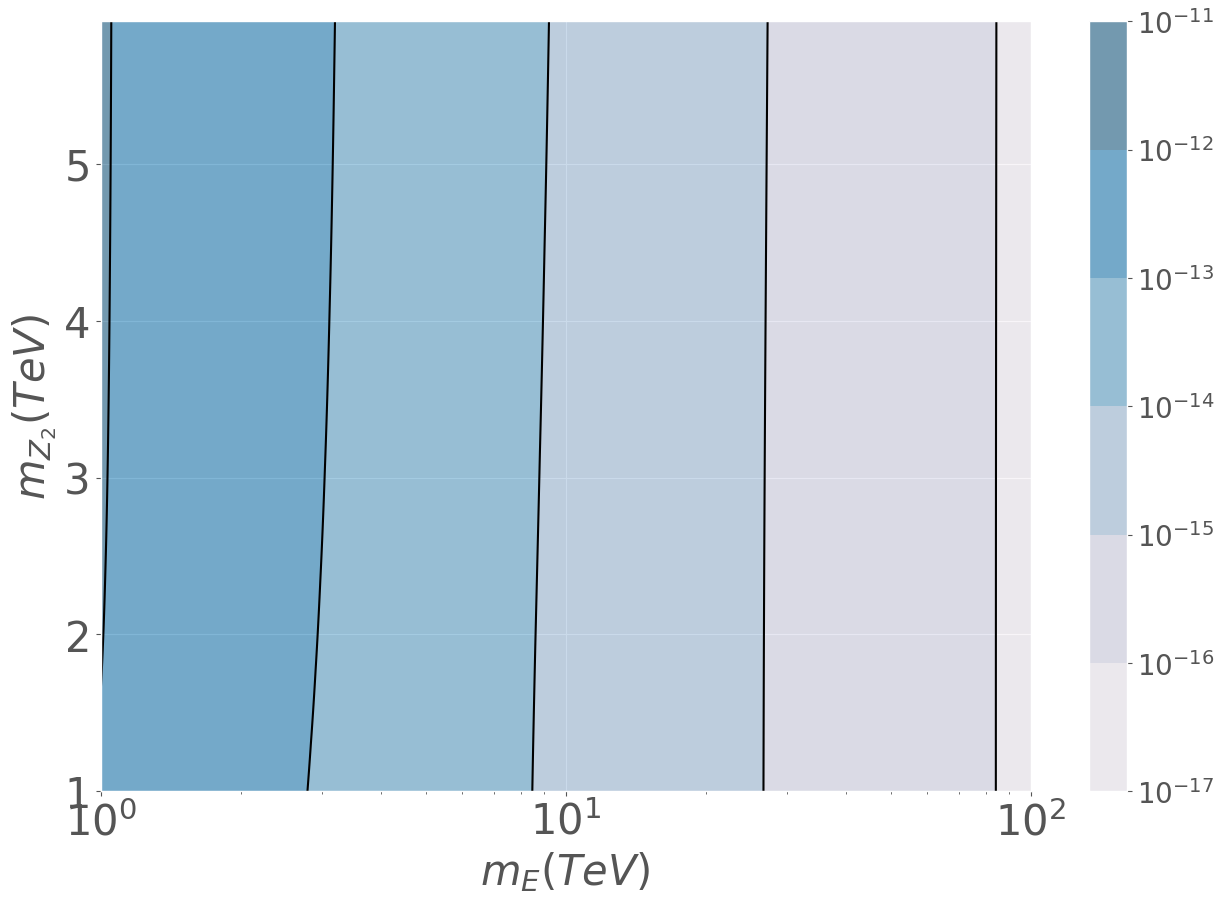}
    \caption{ Contours of the absolute value $|\Delta a_{\mu}^{Z_{2}}|$ of muon $g-2$ contribution due to a $Z_{2}$ gauge boson and  an exotic lepton $E$ in the inner loop as a function of their masses for $v_{1}=245.6$ GeV, $v_{\chi}=5$ TeV.}
    \label{g-2big}
\end{figure}

After numerical integration, the muon $g-2$ contribution as a function of $m_{E}$ and $m_{Z_{2}}$ is shown in figure \ref{g-2big} for $v_{\chi}=5$ TeV. Since its contribution is negative, the absolute value is shown so it can be logarithmically scaled, meaning that it cannot explain the anomaly by itself. 

\subsection{Flavour changing interactions with \texorpdfstring{$H$}{} and \texorpdfstring{$A^{0}$}{}}
\begin{figure}[ht]
    \centering
    \includegraphics[scale=0.25]{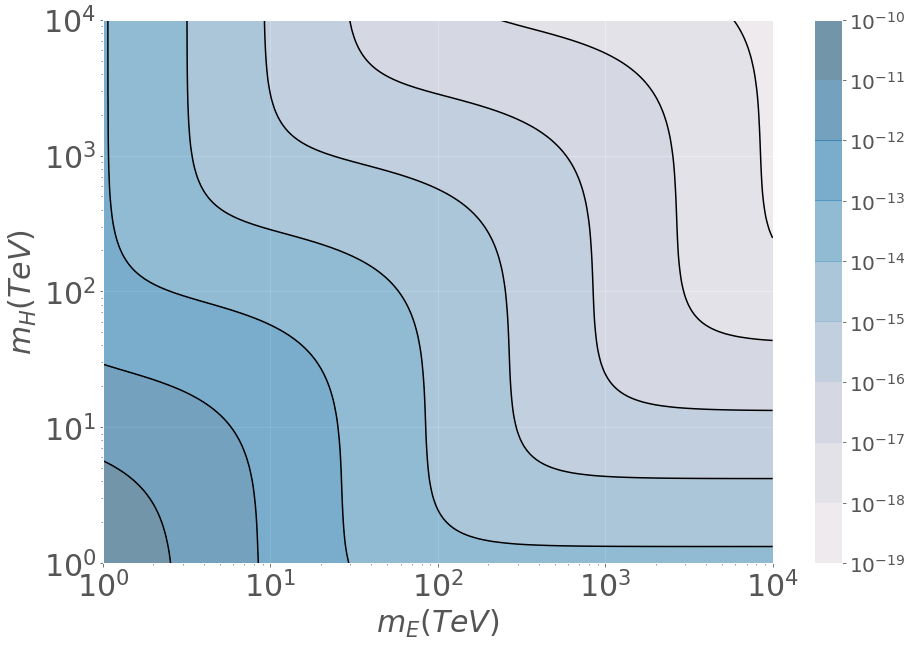}
    \caption{Contours of the absolute value added contributions $|\Delta a_{\mu}^{Z_{2}} + \Delta a_{\mu}^{H} + \Delta a_{\mu}^{A^{0}}|$  to muon $g-2$ due to diagrams in figure \ref{g-2a} and \ref{g-2b} as a function of the charged scalar mass and the exotic lepton mass.}
    \label{g-2small}
\end{figure}

The contributions due to scalars and pseudoscalars are shown in figure \ref{g-2b}. By rotating to mass eigenstates in the Lagrangian in Eq.(\ref{lagE}), we obtain the interaction among muon, scalars and the exotic lepton $E$, which at  order $\mathcal{O}(m_{E}^{-1})$, reads

\begin{align}
    -\mathcal{L}_{\mu\phi E}&= - \frac{m_{\mu}t_{\beta}s_{\beta}}{\sqrt{2}m_{E}}(s_{e\mu} q_{11} + c_{e\mu} q_{12}  ) \bar{E}_{L}\phi\mu_{R} + \frac{c_{\beta}}{\sqrt{2}}(s_{e\mu} q_{11} + c_{e\mu} q_{12}  ) \bar{E}_{R}\phi\mu_{L} ,\label{phileptons}
\end{align}
  \noindent
  where $\phi=(H,iA^{0})$ and using the approximations $c_{13} \approx c_{\gamma} \approx 1$. Then, the contribution to muon $g-2$ can be written as,
  
  \begin{align}
\Delta a_{\mu}^{H,A^{0}} &= \frac{1}{8\pi^2}\frac{m_\mu^2}{ M_{\phi}^2 } \int_0^1 dx \frac{g_{s}^2 \ P_{s}(x) + g_{p}^2 \ P_{p} (x) }{(1-x)(1-\lambda^{2} x) +\epsilon^2 \lambda^2 x},
\end{align}
with
\small
\begin{align}
P_{s}(x) & = x^2(1+\epsilon-x), &
P_{p}(x) & = x^{2}(1-\epsilon -x), \\
\epsilon&= \frac{m_{E}}{m_{\mu}}, & \lambda&=\frac{m_{\mu}}{m_{\phi}},
\end{align}
\normalsize
\noindent
and $g_{s}$ and $g_{p}$ as the scalar and pseudoscalar couplings which can be obtained from Eq.(\ref{phileptons}). Numerical integration shows that this contribution is negative and shows a similar behavior like the $Z_{2}$ contributions which showed large $|\Delta a_{\mu}|$  values for small TeV masses. The total contribution due to scalars, pseudoscalars and $Z_{2}$ is shown in figure \ref{g-2small}, where $m_{H}\approx m_{A^{0}}$ according to section \ref{scalars}. We see that in general there is an important suppression due to $m_{E}$ which can be seen from the couplings depending on $m_{E}^{-1}$, while in the case of the left-handed couplings with scalars, the suppression increases for larges values of $\tan\beta$. In this way, the flavour changing neutral interactions due to $Z_2$, $H$ and $A^0$ provide negligible contributions to muon $g-2$.

\subsection{Charged \texorpdfstring{$W^{+}$}{} boson and exotic neutrinos}

\begin{figure}[ht]
    \centering
    \includegraphics[scale=0.25]{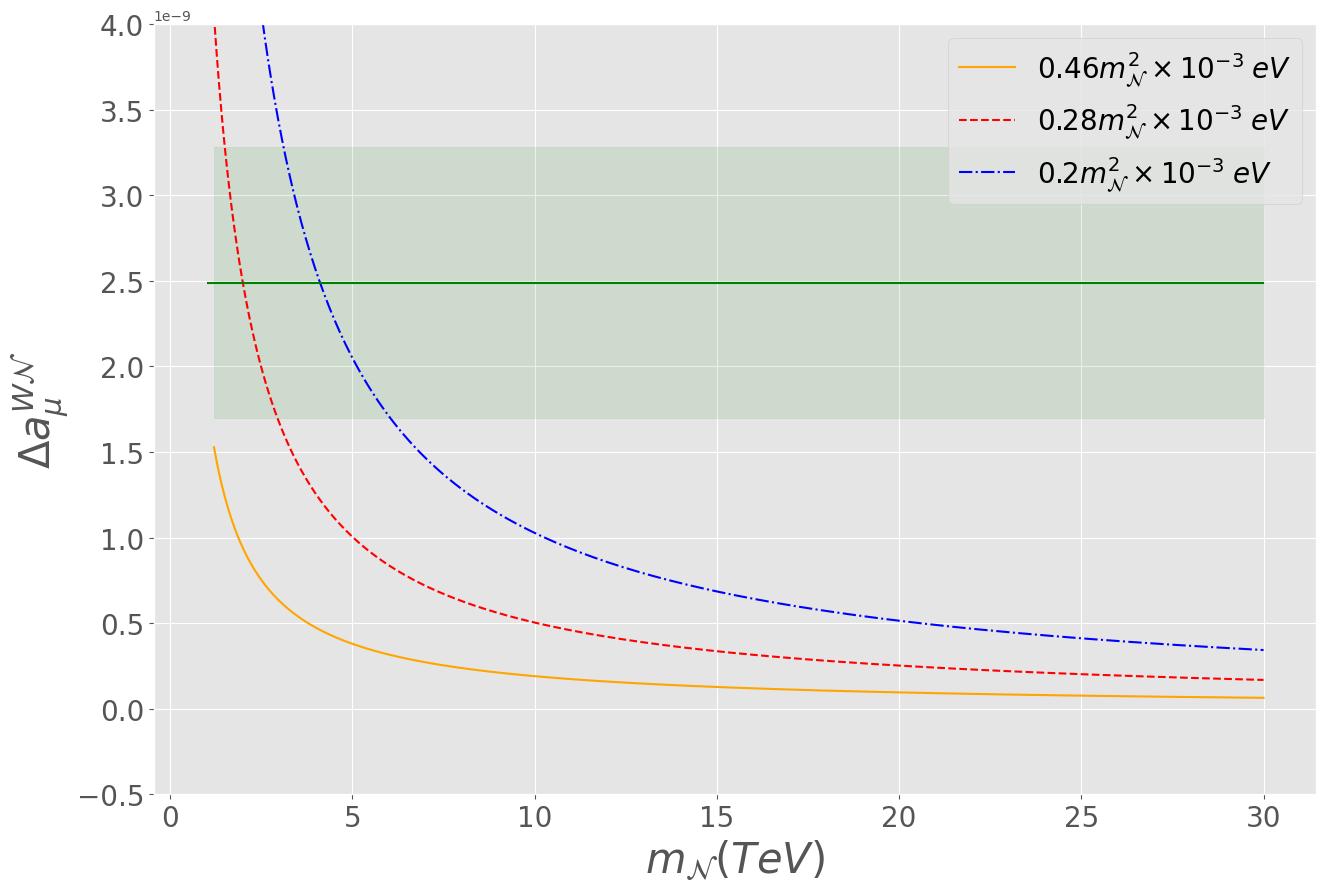}
    \caption{Added contributions $\sum_{i=1}^{6}\Delta a_{\mu}^{W\mathcal{N}_{i}} \equiv \Delta a_{\mu}^{W\mathcal{N}}$  to muon $g-2$ due to a $W^{+}$ and the exotic neutrinos with nearly degenerate masses as a function of $m_{\mathcal{N}}$ for different values of $\mu_{N}$. The green region represents the experimental value at 90\% C.L..}
    \label{g-2W}
\end{figure}

From the electroweak charged current, we can obtain an interaction involving the muon, $W^{+}$ and exotic neutrinos $\mathcal{N}$. The interaction in terms of mass eigenstates can be written as,
\begin{align}
    \mathcal{L}_{W\mathcal{N}\mu}&=- \frac{g  v_{2} \gamma^{\mu}}{4m_{\mathcal{N}}} \left[ s_{\theta_{e\mu}} h_{2e}(j) +c_{\theta_{e\mu}}h_{2\mu}(j) \right]\bar{\mathcal{N}}^{j} W_{\mu}^{+} P_{L}\mu ,
\end{align}
\noindent
where the $h_{2e,\mu}(j)$ couplings are defined by,
\begin{align}
   h_{2e,\mu}(1) &= -i h_{2e,\mu}^{\nu e}, & h_{2e,\mu}(4) &= h_{2e,\mu}^{\nu e}, \nonumber \\
   h_{2e,\mu}(2) &= -i h_{2e,\mu}^{\nu \mu}, & h_{2e,\mu}(5) &= h_{2e,\mu}^{\nu \mu}, \nonumber \\
   h_{2e,\mu}(3) &= -ih_{2e,\mu}^{\nu \tau}, & h_{2e,\mu}(6) &= h_{2e,\mu}^{\nu \tau},    \label{hemudefs}
\end{align}
\noindent
and $j=1,...,6$. Their contribution to muon $g-2$ is according to the loop diagram shown in figure \ref{g-2c} which can be written as,

 \begin{align}
\Delta a_{\mu}^{W\mathcal{N}_{j}} &= \frac{1}{8\pi^2}\frac{m_\mu^2}{ m_{W}^2 } \int_0^1 dx \frac{g_{s}^2 \ P_{s}(x) + g_{p}^2 \ P_{p} (x) }{(1-x)(1-\lambda^{2} x) +\epsilon^2 \lambda^2 x},
\end{align}
where 
\small
\begin{align}
P_{s}(x) & = 2x^{2}(1+x-2\epsilon) + \lambda^{2}(1-\epsilon)^{2}x(1-x)(x+\epsilon), \nonumber\\
P_{p}(x) & = 2x^{2}(1+x+2\epsilon) + \lambda^{2}(1+\epsilon)^{2}x(1-x)(x-\epsilon), 
\end{align}
\normalsize
\noindent
$\epsilon= \frac{m_{\mathcal{N}}}{m_{\mu}}$ and  $\lambda=\frac{m_{\mu}}{m_{W}}$. This contribution is positive and highly sensitive to neutrino Yukawa couplings $h_{2e}^{\nu q}$ and $h_{2\mu}^{\nu q}$.
Massive neutrinos have a mass around the $10^{-3}$ eV scale, whose smallness can be justified by the overall factor $\frac{\mu_{N}v_{2}^{2}}{h_{N_{\chi 1}}^{2}v_{\chi}^{2}}$, as shown in appendix \ref{neutrinomasses} . In this way, we see that the smaller the factor is, the larger the Yukawa couplings are. Such requirement translate in an estimate for the $\mu_{N}$ parameter given by,
\begin{align}
    \frac{\mu_{N}v_{2}^{2}}{h_{N_{\chi 1}}^{2}v_{\chi}^{2}} &= \frac{\mu_{N}v_{2}^{2}}{2 m_{\mathcal{N}}^{2}}\sim 10^{-3} \text{ eV}.
\end{align}
The plot in figure \ref{g-2W} shows the behavior of the $g-2$ contribution as a function of the exotic neutrino masses for different values of $\mu_{N}$. Since a lower bound for heavy Majorana neutrinos of $1.2$ TeV was reported in \cite{Nbound}, we can obtain an upper bound on $\mu_{N}$ according to such mass and the muon $g-2$ at $90\%$ C.L.  given by $\mu_{N}=0.45 m_{\mathcal{N}}^{2} \times 10^{-3}$ eV (orange curve). Nevertheless, for smaller values of $\mu_{N}$ we obtain larger contributions for relative small exotic neutrino masses.

\subsection{Charged scalars and exotic neutrinos}
\begin{figure}[ht]
    \centering
    \includegraphics[scale=0.25]{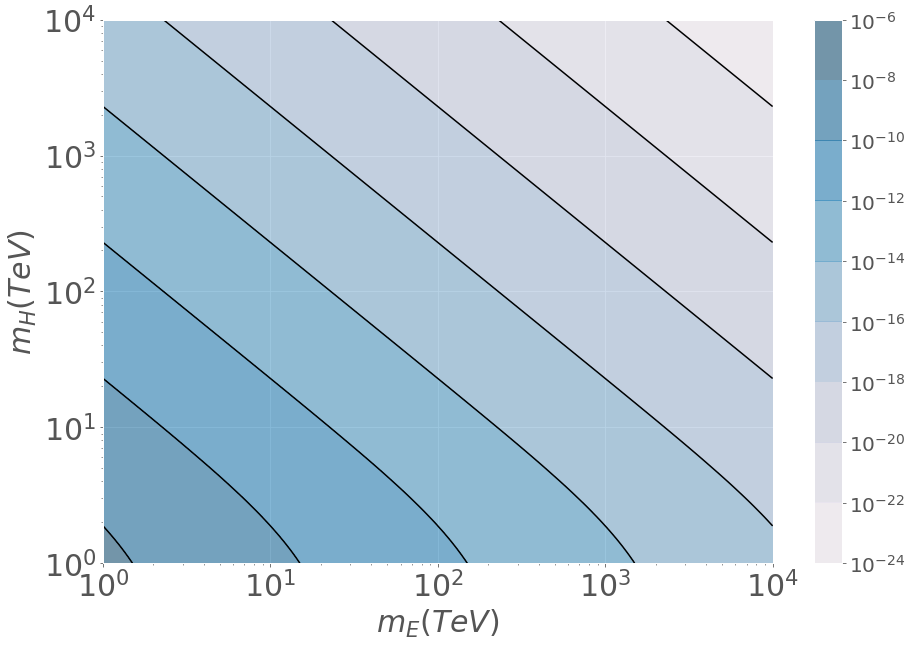}
    \caption{Contours of the absolute value added contributions $|\sum_{i=1}^{6} \Delta a_{\mu}^{H^{\pm}\mathcal{N}_{j}}| \equiv |\Delta a_{\mu}^{H^{\pm}\mathcal{N}}|$ to muon $g-2$ due to the six nearly degenerate neutrinos as a function of the charged scalar mass $m_{\boldsymbol{H}^{\pm}} \approx m_{H}$ and the exotic lepton mass for $m_{\mathcal{N}}=1.2$ TeV.}
    \label{g-2HC}
\end{figure}

An important contribution comes by considering charged scalars $H^{\pm}$ and exotic neutrinos $\mathcal{N}_{j}$ as shown in figure \ref{g-2d}. By rotating to mass eigenstates, we obtain from the Lagrangian in Eq. (\ref{lagN}) the relevant interactions among heavy neutrinos, charged scalars and the muon,

\begin{align}
    \mathcal{L}_{\mathcal{N}H^{\pm}\mu} & =\frac{v_{1}s_{\beta}}{\sqrt{2}m_{E}}\left[q_{11}h_{2e}^{*}(q)+q_{12}h_{2\mu}^{*}(q)\right](R_{\nu}^{\dagger})_{kq}\bar{\nu}_{R}^{k}H^{+}E_{L},
\end{align}

\noindent
where $k=4,...,9$ is used only for exotic neutrino mass eigenstates, $q=4,5,6$ and $h_{2e,\mu}(q)$ is defined in Eq.(\ref{hemudefs}). The muon $g-2$ contribution can be written as,

  \begin{align}
\Delta a_{\mu}^{H^{\pm}\mathcal{N}_{j}} &= \frac{1}{8\pi^2}\frac{m_\mu^2}{ M_{H^{+}}^2 } \int_0^1 dx \frac{g_{s}^2 \ P_{s}(x) + g_{p}^2 \ P_{p} (x) }{(1-x)(1-\lambda^{2} x) +\epsilon^2 \lambda^2 x},
\end{align}
where 
\small
\begin{align}
P_{s}(x) & = -x(1-x)(x+\epsilon), &
P_{p}(x) & = -x(1-x)(x-\epsilon), \\
\epsilon&= \frac{m_{\mathcal{N}}}{m_{\mu}}, & \lambda&=\frac{m_{\mu}}{m_{H^{+}}}.
\end{align}
\normalsize
Since $m_{\boldsymbol{H}^{\pm}} \approx m_{H}$ we can compare this contribution to the neutral scalar case by considering nearly degenerate exotic neutrinos as well, so we can add the contributions due to all six neutrinos, which gives the absolute value for the muon $g-2$ shown in figure \ref{g-2HC}. Despite having a negative contribution, it provides larger values than the neutral scalar ones, which all together compensate the $W^{+}$  gauge boson contribution to fit the anomaly.

\subsection{Total \texorpdfstring{$g-2$}{} prediction}
\begin{figure}[h]
    \centering
    \includegraphics[scale=0.25]{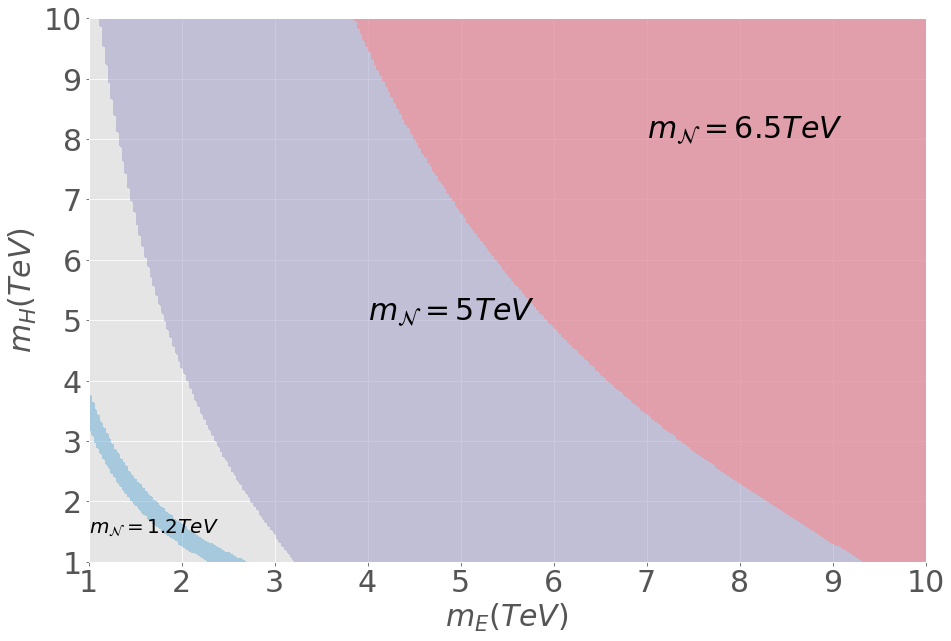}
    \caption{Allowed masses for the exotic lepton $E$ and heavy scalars compatible with muon $g-2$ at 90 \% C.L. for different exotic neutrino masses, $m_{\mathcal{N}}=1.2$ TeV in blue, $m_{\mathcal{N}}=5$ TeV in purple and $m_{\mathcal{N}}=12$ TeV in pink. The total $g-2$ contribution has been calculated as $|\Delta a_{\mu}^{Tot}| = | \Delta a_{\mu}^{Z_{2}} + \Delta a_{\mu}^{W\mathcal{N}} + \Delta a_{\mu}^{H^{\pm}\mathcal{N}} + \Delta a_{\mu}^{H} + \Delta a_{\mu}^{A^{0}}|$.}
    \label{g-2Tot}
\end{figure}

We saw that the interaction with $W^{+}$ bosons provides a positive contribution to muon $g-2$ while in the case of neutral and charged scalars, the contributions are negative with large values for small masses in the TeV scale i.e. $1-2$ TeV. Now, we add all contributions shown in figure \ref{g-2feynman} to find the allowed region in the parameter space fitting the anomaly. In particular, we consider $\mu_{N}=0.2 m_{\mathcal{N}}^{2} \times 10^{-3}$ eV, nearly degenerate exotic neutrino masses and the parameter choice shown in Eq. (\ref{params}). In figure \ref{g-2Tot} we display the allowed regions compatible with the muon $g-2$ for three different values of $m_{\mathcal{N}}$. \\

First, the lower mass bound for exotic neutrinos is taken at $1.2$ TeV (blue region), according to the ATLAS experiment \cite{Nbound}. In this case, the positive contribution due to diagram \ref{g-2c} is larger than the experimental value, so small $m_{E}$ and $m_{H}$ masses are required to generate a negative contribution of the same order that counteracts its value. Furthermore, from figure \ref{g-2W} we see that there is an upper bound for the exotic neutrino mass of $6.7$ TeV at 90\% C.L. from muon $g-2$, where the anomaly can be explained entirely by the interaction with $W$ bosons (diagram $\ref{g-2c}$) and contributions due to heavy scalars must be small enough to not decrease the total $g-2$ out of the 90 \% C.L. interval. For instance,  when the heavy neutrino takes a mass value close to the upper limit of $6.5$ TeV, represented by the pink region in figure \ref{g-2Tot}, a lower mass bound for $m_{E}$ and $m_{H}$ is represented by the boundary of the region, whose values are larger than the bounds set for intermediate values of $m_{\mathcal{N}}$, such as $m_{\mathcal{N}}=5$ TeV in the purple region which lead to larger negative contributions. It is also worth noting that the total $g-2$ implies masses of the same order, so in spite of the $v_{\chi}$ dependence of $m_{E}$, $m_{\mathcal{N}}$ and $m_{H}$, they can be justified by Yukawa couplings of order 1.

\section{\texorpdfstring{$B$}{} meson anomalies}\label{Banom}
The $B$ meson neutral anomalies are dominated by the 2020 LHCb data \cite{LHCb:2020lmf} and could at first be understood in the present model at the LO\footnote{We note that one-loop level penguin and box contributions involving both the heavy fermions and the $W$ boson can appear and be further correlated to the parameter space of the anomalous muon $g-2$ solution. In the present work, we however restrict the study at the LO level, leaving a more comprehensive analysis addressing the NLO effects for a future follow-up work.} by considering flavour changing neutral interactions mediated by $Z_{1}$ and $Z_{2}$ (see Appendix \ref{gaugebosons}), according to the diagram shown in figure \ref{BanomDiag}.

\begin{figure}[ht]
    \centering
    \huge
    \begin{tikzpicture}
  \begin{feynman}
    \vertex (a) {\(b\)};
    \vertex [above = of a](a1) {\(d\)};
    \vertex [right=of a] (b);
    \vertex [right=of b] (f1);
    \vertex [right=of a1] (b1);
    \vertex [right=of b1] (f11);
    \vertex [ right=of f1] (j) {\(\overline{s}\)};
    \vertex [ right=of f11] (j1) {\(d\)};
    \vertex [below right=of b] (c);
    \vertex [right=of c] (f2) {\(l^+\)};
    \vertex [below right=of c] (f3) {\(l^{-}\)};

    \diagram* {
      (a) -- [fermion] (b) -- [plain] (f1)-- [fermion] (j),
      (a1) -- [fermion] (b1) -- [plain] (f11)-- [ fermion ] (j1),
      (b) -- [boson, edge label'=\(Z_{1,2}\)] (c),
      (c) -- [anti fermion] (f2),
      (c) -- [fermion] (f3),
    };
  \end{feynman}
\end{tikzpicture}
    \caption{Decay $B^{0}\rightarrow K^{*0}\ell^{+}\ell^{-}$ due to neutral gauge bosons $Z_{1}$ and $Z_{2}$.}
    \label{BanomDiag}
\end{figure}

\subsection{Neutral current fermion couplings}
The relevant neutral current interactions can be written as,
\begin{align}
    \mathcal{L}_{ij}=ig\bar{f}_{Li}[ J_{i}^{L1}\slashed{Z}_{1} &+ J_{i}^{L2}\slashed{Z}_{2}]f_{Li} + ig\bar{f}_{Ri}[ J_{i}^{R1}\slashed{Z}_{1} + J_{i}^{R2}\slashed{Z}_{2}]f_{Ri},
\end{align}
\noindent
where $f_{i}$ runs over all fermions in the flavour basis and the $J_{i}^{L1,2}$, $J_{i}^{R1,2}$ couplings can be read off from tables \ref{JL} and \ref{JR}.

\begin{table}[ht]
    \centering
    \begin{tabular}{|p{0.03\textwidth}|p{0.22\textwidth}|p{0.23\textwidth}|} \hline
       $ f_{Li}      $ & $ J_{i}^{L1} $ & $ J_{i}^{L2}$ \\\hline
       $ u_{L}^{1}   $ & $ \frac{1}{c_{W}}\left(-\frac{1}{2}+ \frac{2}{3}s_{W}^{2}\right)c_{Z} - \frac{g_{X}}{3g}s_{Z} $ & $- \frac{1}{c_{W}}\left(-\frac{1}{2}+ \frac{2}{3}s_{W}^{2}\right)s_{Z} - \frac{g_{X}}{3g}c_{Z} $ \\\hline
        $u_{L}^{2,3} $ & $ \frac{1}{c_{W}}\left(-\frac{1}{2}+ \frac{2}{3}s_{W}^{2}\right)c_{Z} $ & $ -\frac{1}{c_{W}}\left(-\frac{1}{2}+ \frac{2}{3}s_{W}^{2}\right)s_{Z} $ \\\hline
        $d_{L}^{1} $ & $ \frac{1}{c_{W}}\left( \frac{1}{2} - \frac{1}{3}s_{W}^{2}\right)c_{Z} - \frac{g_{X}}{3g}s_{Z} $ & $ -\frac{1}{c_{W}}\left( \frac{1}{2} - \frac{1}{3}s_{W}^{2}\right)s_{Z} - \frac{g_{X}}{3g}c_{Z} $ \\\hline
        $d_{L}^{2,3}$ & $ \frac{1}{c_{W}}\left(\frac{1}{2} - \frac{1}{3}s_{W}^{2}\right)c_{Z} $ & $ -\frac{1}{c_{W}}\left(\frac{1}{2} - \frac{1}{3}s_{W}^{2}\right)s_{Z} $ \\\hline
        $e_{L}^{e,\mu} $ & $ \frac{1}{c_{W}}\left(\frac{1}{2} - s_{W}^{2}\right)c_{Z} $ & $ -\frac{1}{c_{W}}\left(\frac{1}{2} - s_{W}^{2}\right)s_{Z} $ \\ \hline
        $e_{L}^{\tau} $ & $ \frac{1}{c_{W}}\left(\frac{1}{2} - s_{W}^{2}\right)c_{Z} + \frac{g_{X}}{g}s_{Z} $ & $- \frac{1}{c_{W}}\left(\frac{1}{2} - s_{W}^{2}\right)s_{Z} + \frac{g_{X}}{g}c_{Z} $ \\\hline
        $E_{L} $ & $ -\frac{s_{W}^{2}}{c_{W}}c_{Z} + \frac{g_{X}}{g}s_{Z} $ & $ \frac{s_{W}^{2}}{c_{W}}s_{Z} + \frac{g_{X}}{g}c_{Z} $ \\\hline
        $\mathcal{T}_{L} $ & $ \frac{2s_{W}^{2}}{3c_{W}}c_{Z} - \frac{g_{X}}{3g}s_{Z} $ & $ -\frac{2s_{W}^{2}}{3c_{W}}s_{Z} - \frac{g_{X}}{3g}c_{Z} $ \\\hline
        $\mathcal{J}_{L}^{a} $ & $ -\frac{s_{W}^{2}}{3c_{W}}c_{Z} $ & $ \frac{s_{W}^{2}}{3c_{W}}s_{Z} $ \\\hline
    \end{tabular}
    \caption{Neutral current couplings for left-handed fermions.}
    \label{JL}
\end{table}
\normalsize
\begin{table}[ht]
    \centering
    \begin{tabular}{|p{0.07\textwidth}|p{0.2\textwidth}|p{0.21\textwidth}|} \hline
     $f_{Ri}$       &  $J_{i}^{R1}$  &  $J_{i}^{R2}$ \\ \hline
     $U_{R}^{1,2,3}    $ & $  \frac{2s_{W}^{2}}{3c_{W}}c_{Z} - \frac{2g_{X}}{3g}s_{Z} $ & $  -\frac{2s_{W}^{2}}{3c_{W}}s_{Z} - \frac{2g_{X}}{3g}c_{Z} $ \\ \hline
      $D_{R}^{1,2,3}  $ & $  -\frac{s_{W}^{2}}{3c_{W}}c_{Z} + \frac{g_{X}}{3g}s_{Z}  $ & $  \frac{s_{W}^{2}}{3c_{W}}s_{Z} + \frac{g_{X}}{3g}c_{Z}  $ \\ \hline
      $e_{R}^{e,\tau} $ & $  -\frac{s_{W}^{2}}{c_{W}}c_{Z} + \frac{4g_{X}}{3g}s_{Z} $ & $   \frac{s_{W}^{2}}{c_{W}}s_{Z} + \frac{4g_{X}}{3g}c_{Z} $ \\ \hline
      $e_{R}^{\mu} $ & $  -\frac{s_{W}^{2}}{c_{W}}c_{Z} + \frac{g_{X}}{3g}s_{Z} $ & $  \frac{s_{W}^{2}}{c_{W}}s_{Z} + \frac{g_{X}}{3g}c_{Z} $ \\ \hline
      $E_{R} $ & $  -\frac{s_{W}^{2}}{c_{W}}c_{Z} + \frac{2g_{X}}{3g}s_{Z}  $ & $  \frac{s_{W}^{2}}{c_{W}}s_{Z} + \frac{2g_{X}}{3g}c_{Z} $ \\ \hline
      $\mathcal{T}_{R} $ & $  \frac{2s_{W}^{2}}{3c_{W}}c_{Z} - \frac{2g_{X}}{3g}s_{Z} $ & $  -\frac{2s_{W}^{2}}{3c_{W}}s_{Z} - \frac{2g_{X}}{3g}c_{Z} $ \\ \hline
      $\mathcal{J}_{R}^{a} $ & $  -\frac{s_{W}^{2}}{3c_{W}}c_{Z} + \frac{g_{X}}{3g}s_{Z} $ & $   \frac{s_{W}^{2}}{3c_{W}}s_{Z} + \frac{g_{X}}{3g}c_{Z} $ \\ \hline
    \end{tabular}
    \caption{Neutral current couplings for right-handed fermions.}
    \label{JR}
\end{table}
\noindent
In general, the $Z_{2}^{\mu}$ couplings can be obtained by exchanging the sign in the electroweak term of $Z_{1}^{\mu}$ (first term) and by doing the replacement $s_{Z} \leftrightarrow c_{Z}$. Rotating the fermions into mass eigenstates $F_{m}$, the interaction Lagrangian becomes,

\begin{align}
    \mathcal{L}_{mn}=&ig\bar{F}_{Lm}(\mathbb{V}_{L}^{\dagger})^{mi}[ J_{i}^{L1}\slashed{Z}_{1} + J_{i}^{L2}\slashed{Z}_{2}](\mathbb{V}_{L})^{in}F_{Ln}  + \bar{F}_{Rm}(\mathbb{V}_{R}^{\dagger})^{mi}[ J_{i}^{R1}\slashed{Z}_{1} + J_{i}^{R2}\slashed{Z}_{2}](\mathbb{V}_{R})^{in}F_{Rn}. \label{lagNC}
\end{align}

From the Lagrangian in Eq.(\ref{lagNC}), we take  $m=2$ and $n=3$ for the down quark sector to extract the down-strange flavour changing interaction. First, from table \ref{JR} we see that there is right-handed down-like universality, so its contribution to the flavour changing Lagrangian vanishes due to the unitarity of the rotation matrix, $(\mathbb{V}_{R}^{D\dagger})^{mi}(\mathbb{V}_{R}^{D})^{in}=\delta_{mn}$. Secondly, for the left-handed particles we can split the couplings into both SM and exotic terms as,

\begin{align}
 (\mathbb{V}_{L}^{D \dagger})^{2i} &J_{i}^{L l}(\mathbb{V}_{L}^{D})^{i3} =  (\mathbb{V}_{L}^{D \dagger})^{21} J_{1}^{Ll}(\mathbb{V}_{L}^{D})^{13} + (\mathbb{V}_{L}^{D \dagger})^{2r} J_{r}^{Ll}(\mathbb{V}_{L}^{D})^{r3} +  (\mathbb{V}_{L}^{D \dagger})^{2\alpha} J_{\alpha}^{Ll}(\mathbb{V}_{L}^{D})^{\alpha 3},
\end{align}
 \noindent
where $r=2,3$ labels the second and third generation of quarks, $\alpha=4,5$ labels the exotic quarks and $l=1,2$ labels the neutral gauge bosons. Then, from table \ref{JL}, we can see that $J_{2}^{L1,2}=J_{3}^{L1,2}$. Thus, we can use the unitarity constraint to replace,
\begin{align}
    (\mathbb{V}_{L}^{D\dagger})^{2r}(\mathbb{V}_{L}^{D})^{r3} &= -(\mathbb{V}_{L}^{D\dagger})^{21}(\mathbb{V}_{L}^{D})^{13} -(\mathbb{V}_{L}^{D\dagger})^{2\alpha}(\mathbb{V}_{L}^{D})^{\alpha},
\end{align}
in such a way that the interaction Lagrangian can be written as,
 
 \begin{align}
     \mathcal{L}_{bs}&=\bar{s}[ g(\mathbb{V}_{L}^{D\dagger})^{21}(J_{1}^{Ll}-J_{r}^{L1})(\mathbb{V}_{L}^{D})^{13} \slashed{Z}_{l}]P_{L}b +\bar{s}[ g(\mathbb{V}_{L}^{D\dagger})^{2\alpha}(J_{\alpha}^{L2}-J_{r}^{L1})(\mathbb{V}_{L}^{D})^{\alpha 3} \slashed{Z}_{l}]P_{L}b \nonumber \\
         &=\bar{s} g(\mathbb{V}_{L}^{D\dagger})^{21}\left(-\frac{1}{3}\frac{g_{X}}{g} (s_{Z}\delta_{l1} + c_{Z}\delta_{l2})\right)(\mathbb{V}_{L}^{D})^{13} \slashed{Z}_{l}P_{L}b +\bar{s} g(\mathbb{V}_{L}^{D\dagger})^{2\alpha}\left(-\frac{1}{2}\frac{s_{Z}}{g}\right)(\mathbb{V}_{L}^{D})^{\alpha 3} \slashed{Z}_{l}P_{L}b.
 \end{align}

It can be seen that the second term is proportional to $s_{Z}$ for both $Z_{1}$ and $Z_{2}$ which initially suppresses the contribution as indicated by LEP data \cite{ALEPH:2005ab,Langacker:2009su}. Additionally, working in the decoupling limit from now on, we can see that the entries $(\mathbb{V}_{L}^{D})^{\alpha3}$
 and $(\mathbb{V}_{L}^{D\dagger})^{2\alpha}$ of the rotation matrix come from the see-saw decoupling of exotic quarks to SM quarks, so they are proportional to $v_{\chi}/m_{\mathcal{J}^{1}}m_{\mathcal{J}^{2}}$ $\text{GeV}^{-1}$, making them negligible. In this way, the bottom-strange interaction Lagrangian is given by,

\begin{align}
    \mathcal{L}_{bZs}&=-(\mathbb{V}_{L}^{D})_{12}^{*}(\mathbb{V}_{L}^{D})_{13}\left( \frac{g_{X}}{3}s_{Z} \right) \bar{s}\slashed{Z}_{1}P_{L}b -(\mathbb{V}_{L}^{D})_{12}^{*}(\mathbb{V}_{L}^{D})_{13}\left( \frac{g_{X}}{3}c_{Z} \right) \bar{s}\slashed{Z}_{2}P_{L}b  \nonumber\\
    &\approx  g^{bs}_{L}\bar{s}\slashed{Z}_{2}P_{L}b\label{eq:Zbs},
\end{align}
with 
\begin{equation}
    g^{bs}_{L}\equiv-\frac{g_{X}}{3}(\mathbb{V}_{L}^{D})_{12}^{*}(\mathbb{V}_{L}^{D})_{13},\label{eq:bZs}
\end{equation}

\noindent and the left-handed rotations for the down-like quarks can be written as,

\begin{align}
    (\mathbb{V}_{L}^{D})_{13}& = r_{1}^{D}\approx \mathbb{V}_{13}c_{uc}+ \mathbb{V}_{23}s_{uc} &
    (\mathbb{V}_{L}^{D})_{12}^{*} &=\sin\theta_{ds}= s_{ds}\approx\mathbb{V}_{12}^{*}c_{uc}+\mathbb{V}_{22}^{*}s_{uc},
\end{align}
\noindent
where $\mathbb{V}$ is the CKM matrix,  $r_{1}^{D}$ and $\theta_{ds}$ are defined in Eqs.(\ref{r1D}) and (\ref{thetads}) respectively, while $s_{uc}=\sin\theta_{uc}$ and $c_{uc}=\cos\theta_{uc}$ are defined in Eq.(\ref{thetauc}).

The corresponding couplings for electrons and muons can be obtained in a similar way as for the quarks case. Following  Eq.(66) from \cite{Zpole3} we get,

\begin{align}\mathcal{L}_{Z\ell\ell} & \approx\bar{e}\slashed{Z}_{2}\left[\frac{4g_{X}}{3}P_{R}\right]e+\bar{\mu}\slashed{Z}_{2}\left[\frac{g_{X}}{3}P_{R}+2\,g_{X}\left|(\mathbb{V}_{L}^{E})_{32}\right|^{2}P_{L}\right]\mu\nonumber\\
 &= g_{R}^{ee}\,\bar{e}\slashed{Z}_{2}P_{R}e+g_{R}^{\mu\mu}\,\bar{\mu}\slashed{Z}_{2}P_{R}\mu,\label{Lll}+g_{L}^{\mu\mu}\,\bar{\mu}\slashed{Z}_{2}P_{L}\mu,
\end{align}

where we have defined,

\begin{eqnarray}
g_{R}^{ee} & \equiv & \frac{4g_{X}}{3},\quad g_{R}^{\mu\mu}\equiv\frac{g_{X}}{3},\quad g_{L}^{\mu\mu}\equiv2\,g_{X}\left|(\mathbb{V}_{L}^{E})_{32}\right|^{2}.\label{eq:Zll}
\end{eqnarray}

\subsection{Effective Hamiltonian}

At the bottom quark mass scale, the resulting tree-level effective Hamiltonian from the previous neutral current couplings will be given by, 

\begin{align} 
\mathcal{H}_{\text{eff}}^{NP}&=-\frac{1}{M_{Z_2}^2}\left[\overline{s}\left(g^{bs}_{L}\,P_{L}\right)b\right]\left[\overline{\ell}\gamma ^{\mu}\left(g^{\ell\ell}_{L}\,P_{L}+g^{\ell\ell}_{R}\,P_{R}\right)\ell\right]+\text{h.c.}\label{Heff}
\end{align}
\noindent
where $\ell=e,\mu$ and $M_{Z_{2}} \approx M_{Z'} = g_{X}v_{\chi}/3$ according to Eq.(\ref{mzp}). Such NP operators affects the SM contribution,

\begin{align}
{\cal H}_{\mathrm{eff}}^{\mathrm{SM}} & =-\frac{4G_{F}}{\sqrt{2}}\mathbb{V}_{tb}\mathbb{V}_{ts}^{*}\sum_{i=9,10} C^{\ell\ell}_{\mathrm{SM},i}\mathcal{O}^{\ell\ell}_{i},\label{eq:HeffSM}
\end{align}
where  $\ell=e,\mu$,  $G_F$ is the Fermi constant, $\mathbb{V}_{tb(s)}$ are elements of the CKM matrix and the $C^{\ell\ell}_{i,\mathrm{SM}}$ are the effective SM Wilson coefficients at the scale of the bottom quark mass associated to the operators 
\begin{align}
\mathcal{O}^{\ell\ell}_{9} & =\frac{e^{2}}{16\pi^{2}}(\bar{s}\gamma_{\mu}P_{L}b)(\bar{\ell}\gamma^{\mu}\ell),\qquad\qquad\mathcal{O}^{\ell\ell}_{10}=\frac{e^{2}}{16\pi^{2}}(\bar{s}\gamma_{\mu}P_{L}b)(\bar{\ell}\gamma^{\mu}\gamma_{5}\ell).\label{wilson-operators}
\end{align}

Doing the matching between the effective Hamiltonians in Eqs.(\ref{Heff}) and (\ref{eq:HeffSM}) we obtain the NP Wilson coefficients,

\begin{align}
C_{9}^{\ell\ell} & =-\frac{\pi}{\sqrt{2}G_{F}\alpha \mathbb{V}_{tb}\mathbb{V}_{ts}^{*}}\frac{g_{L}^{bs}(g_{L}^{\ell\ell}+g_{R}^{\ell\ell})}{M_{Z_{2}}^{2}},\label{eq:C9ll}\\
C_{10}^{\ell\ell} & =\frac{\pi}{\sqrt{2}G_{F}\alpha \mathbb{V}_{tb}\mathbb{V}_{ts}^{*}}\frac{g_{L}^{bs}(g_{L}^{\ell\ell}-g_{R}^{\ell\ell})}{M_{Z_{2}}^{2}},\label{eq:C10ll}
\end{align}

being $g_{R}^{ee}$, $g_{R}^{\mu\mu}$ and $g_{L}^{\mu\mu}$ as defined in Eq.(\ref{eq:Zll}), and $g^{bs}_L$ from Eq.(\ref{eq:bZs}).

\subsection{Fit to \texorpdfstring{$B$}{} meson decays}

We make use of the \textsf{flavio} package \cite{Straub:2018kue} with the same observables as in \cite{Greljo:2022jac} including the latest LHCb measurements for $R_{K^{(*)}}$ \cite{LHCb:2022qnv,LHCb:2022vje} and the latest CMS-ATLAS-LHCb combination of $\mathrm{BR}(B_{s}\rightarrow\mu^{+}\mu^{-})$ presented in \cite{Greljo:2022jac}. In particular, 96 of those observables are individual bins reported by LHCb \cite{LHCb:2020lmf}, CMS \cite{CMS:2017ivg} and ATLAS \cite{ATLAS:2018gqc} related to the coefficients in the angular distributions of $B^0\to K^{*0}\mu^+\mu^-$ and $B^+\to K^{*+}\mu^+\mu^-$ decays \cite{Ali:1999mm,Hiller:2003js,Bobeth:2007dw,Altmannshofer:2008dz,Descotes-Genon:2013vna}.  \\

We define the model-independent quadratic approximation to the likelihood function 
\begin{equation}
\log\mathcal{L}=-\frac{\chi^{2}}{2},\quad\chi^{2}(\mathbf{C})\approx\chi_{min}^{2}+\frac{1}{2}\left(\mathbf{C}-\mathbf{C}_{\mathrm{bf}}\right)^{T}\mathrm{Cov}^{-1}\left(\mathbf{C}-\mathbf{C}_{\mathrm{bf}}\right),
\end{equation}
where $\mathbf{C}_{\mathrm{bf}}$ is a vector with components defined by the best fit values of the WCs obtained by \textsf{flavio},

\begin{equation}
C_{9,\mathrm{bf}}^{\mu\mu}=-0.56,\:C_{10,\mathrm{bf}}^{\mu\mu}=-0.06,\:C_{9,\mathrm{bf}}^{ee}=C_{10,\mathrm{bf}}^{ee}=0.28,\:\textrm{Pull}_{\textrm{SM}}=4.3\sigma,
\end{equation}

with the $\textrm{Pull}_{\textrm{SM}}$ metric calculated as in \cite{Capdevila:2017bsm,Capdevila:2018jhy},

\begin{align}
\textrm{Pull}_{\textrm{SM}}=\sqrt{2}\,\text{Erf}^{-1}\left[F(\Delta\chi^2;n_\text{dof})\right],
\label{eq:PullSM}
\end{align}

where $F$ is the $\chi^2$ cumulative distribution function and $n_\text{dof}$ is the number of degrees of freedom. The vector $\mathbf{C}$ is defined in terms of the theoretical Wilson coefficients in Eqs.(\ref{eq:C9ll}-\ref{eq:C10ll}), and $\mathrm{Cov}$ is the covariance matrix or Hessian associated to the correlation matrix $\rho$ given by 

\begin{equation}
\rho=\left(\begin{array}{ccc}
1.00 & 0.61 & 0.06\\
0.61 & 1.00 & 0.09\\
0.06 & 009 & 1.00
\end{array}\right),
\end{equation}

obtained by using the \texttt{MIGRAD} minimization algorithm. With this function at hand, a random generator in \texttt{Mathematica} is requested to find points inside the ellipsoid defined by 
$\Delta\chi^{2}\leq 2\sigma$ 
for 2 degrees of freedom and boundaries defined by,
\begin{equation}
\theta_{uc}\in[0,\,\pi],\qquad \tan\beta\in[1,\,300],\qquad \upsilon_{\chi}\in[1,\,10]\,\mathrm{TeV}.
\end{equation}

\begin{figure}[ht]
\begin{centering}
\includegraphics[scale=0.35]{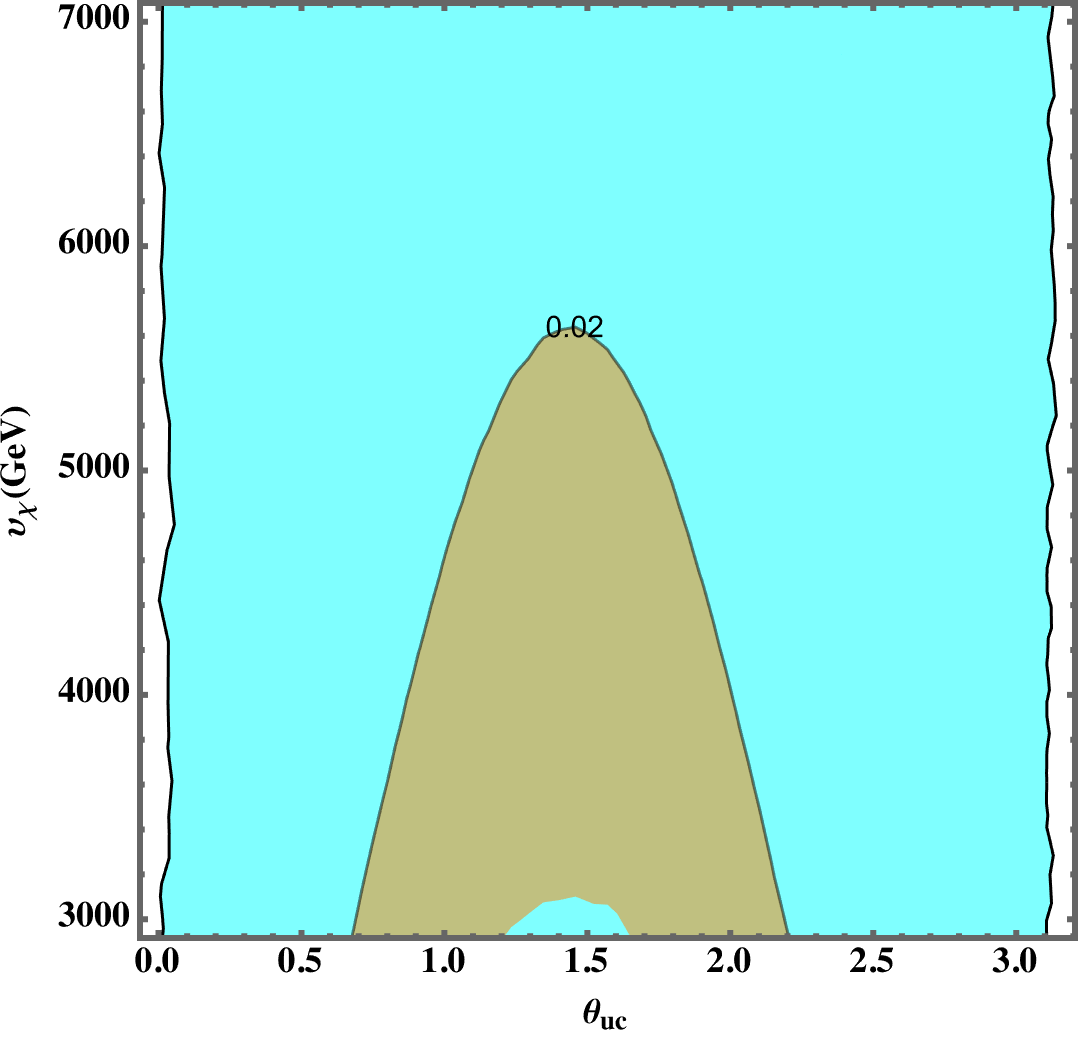} $\qquad\qquad\qquad$\includegraphics[scale=0.35]{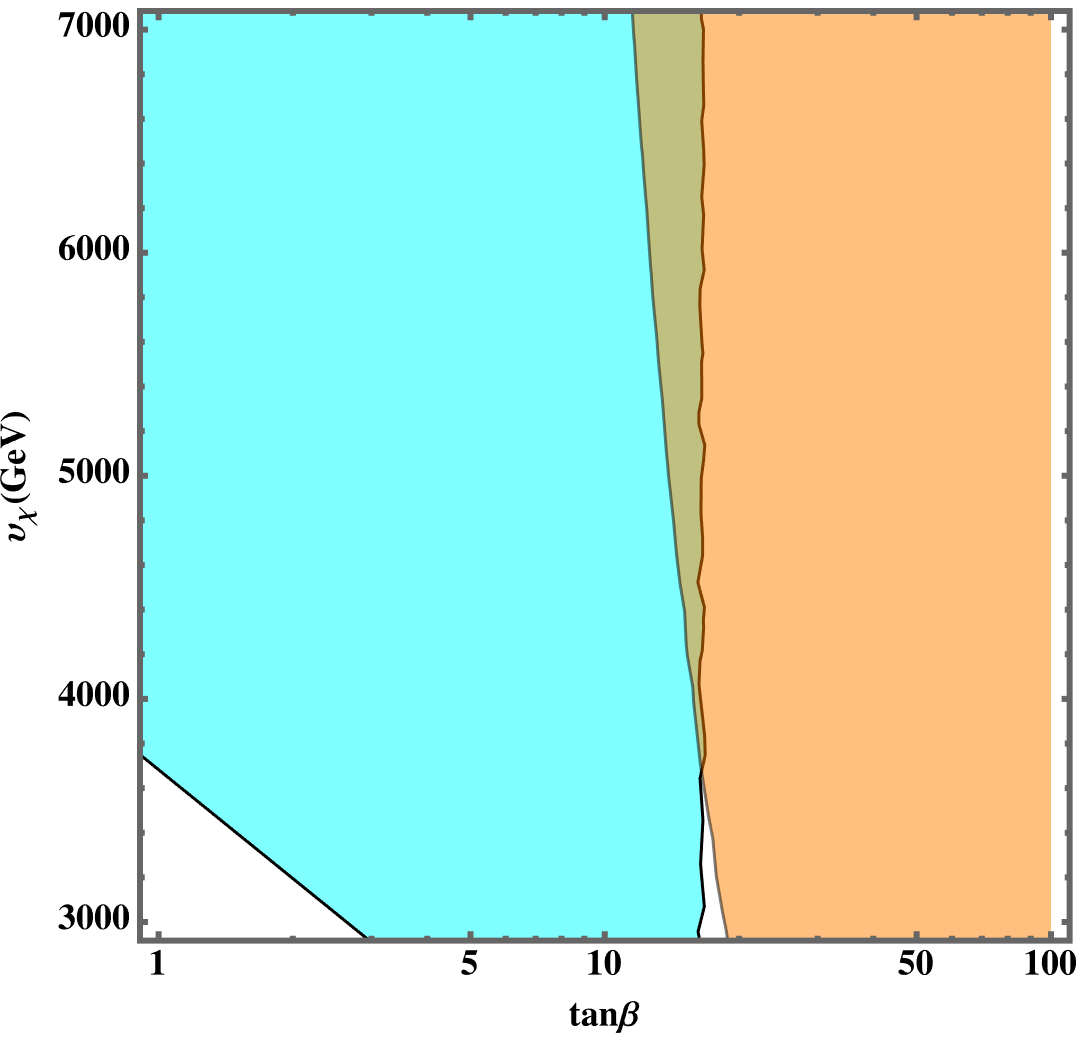} 
\par\end{centering}
\caption{2D Projections of the scanned parameter space and constraints on top. Left: $2\sigma$ region (cyan) in the $\{\upsilon_{\chi},\:\theta_{uc}\}$ plane compatible with the $B$ meson anomalies constrained at $95\%$ C.L. by $\Delta M_{s}$ (allowed region in green). Right: Allowed $95\%$ C.L region by neutrino trident production (green region) in the $\{\upsilon_{\chi},\:\tan\beta\}$ plane. \label{fig:constraints}}
\end{figure}

\begin{figure}[ht]
\begin{centering}
\includegraphics[scale=0.5]{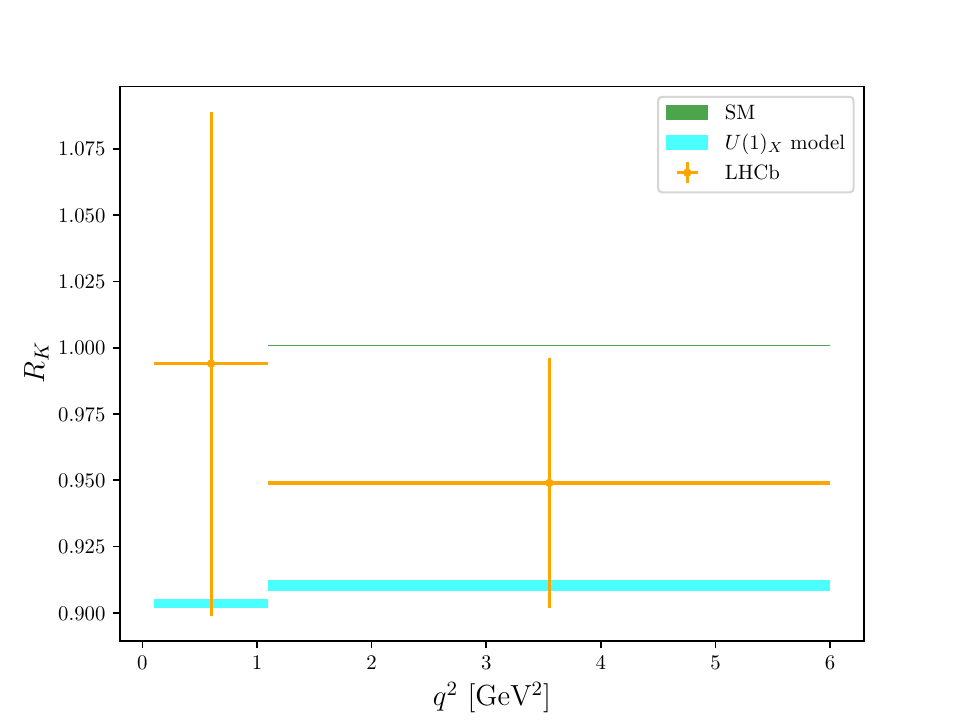} $\qquad$\includegraphics[scale=0.5]{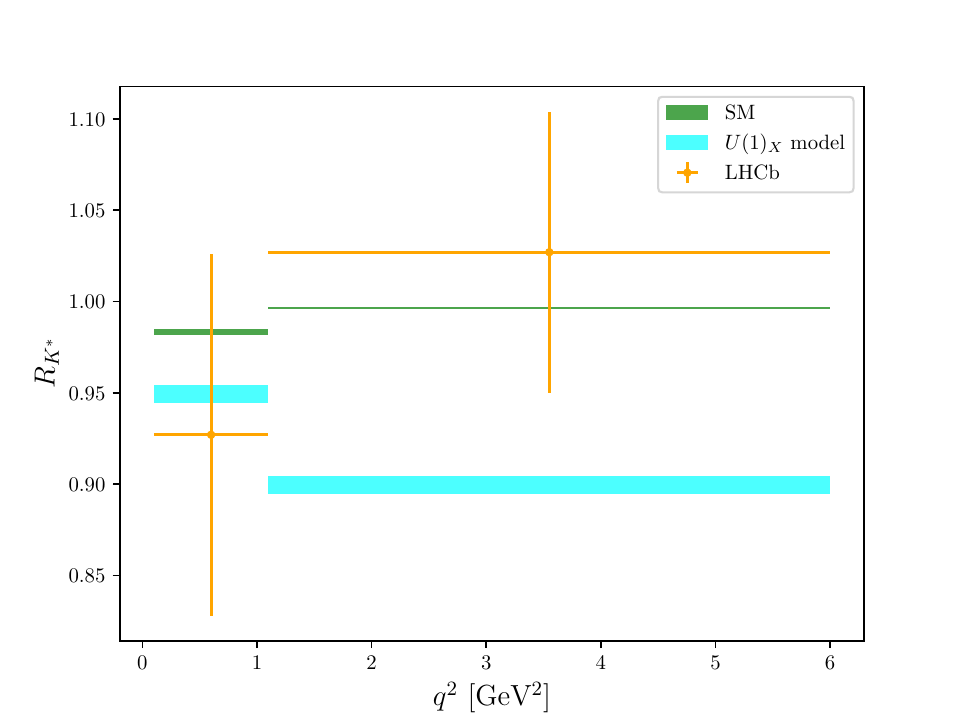} 
\par\end{centering}
\caption{$R_K$ (left) and $R_{K^*}$ (right) bin predictions for the SM (green) and the $U(1)_X$ model (cyan) compared to the LHCb data (orange)\cite{LHCb:2022qnv,LHCb:2022vje}. \label{fig:RK_RK*}}
\end{figure}

We evade more stringent constraints from colliders given that the high-mass Drell-Yan bounds do not apply for lepton flavour violating processes and instead we have to take into account the neutrino trident production cross section \cite{Altmannshofer:2014pba}, given by

\begin{equation}
\frac{\sigma_{\mathrm{SM+NP}}}{\sigma_{\mathrm{SM}}}=1+\frac{\sqrt{2}}{G_{F}}\frac{g_{L}^{\mu\mu}}{M_{Z_{2}}^{2}}\frac{(1+4\sin^{2}\theta_{w})(g_{L}^{\mu\mu}+g_{R}^{\mu\mu})+(g_{L}^{\mu\mu}-g_{R}^{\mu\mu})}{1+(1+4\sin^{2}\theta_{w})^{2}},
\label{eq:trident}\end{equation}

for $\sigma_{\mathrm{exp}}/\sigma_{\mathrm{SM}}=0.83\pm0.18$ \cite{Alguero:2022est}. The resultant parameter space is further constrained on the quark sector by the $\Delta M_{s}$ mass difference from $B_{s}-\overline{B}_{s}$ mixing \cite{DiLuzio:2019jyq,Alguero:2022est,Allanach:2022iod},

\begin{equation}
\left(\frac{g_{L}^{bs}}{0.52}\right)^{2}\left(\frac{10\mathrm{TeV}}{M_{Z_{2}}}\right)^{2}=0.110\pm0.090.\label{eq:DeltaMs}
\end{equation}

In figure \ref{fig:constraints} we show the allowed parameter space from the flavour fit (cyan region) constrained on both the quark and lepton couplings according to Eqs.(\ref{eq:trident}) and (\ref{eq:DeltaMs}). Given that Eq.(\ref{eq:C9ll}) can not generate a positive sign for $C_9^{ee}$, the model is excluded at the $1\sigma$ level as can be seen in figure \ref{fig:RK_RK*} (right) for the intermediate $q^2$ bin, however it can accommodate all the $B$ meson anomalies for $\upsilon_{\chi}\in[4,\,5.7]$ TeV, $\theta_{uc}$ near $\pi/4$ for the largest $\upsilon_{\chi}$ and $\tan\beta\approx15$, favouring $\upsilon_1>\upsilon_2$ as required by the mass hierarchy and further justifying the values used in section \ref{fermions}.

\section{Conclusions}\label{conclusions}
We presented a gauged non-universal $U(1)_X$ extension of the SM in view of the most recent measurements related to the flavour anomalies. We first revisited the model by obtaining general expressions for rotation matrices and mass eigenvalues, where the top quark mass was found to be proportional to $v_{1}$, while the bottom quark, the $\tau$ and muon lepton masses have smaller masses in comparison due to their dependence on $v_{2}$. Moreover, the values of the exotic particle masses are justified by the scalar singlet VEV $v_{\chi}$, which is expected to lie at the TeV scale while the lightest fermions such as the electron and the up, down and strange quarks are massless at tree-level, with their masses being explained by considering the effects of non-renormalizable operators allowed by the $U(1)_{X} \otimes \mathbb{Z}_{2}$ symmetry up to dimension $5$ and $7$, which in general fill out all zeros in the mass matrices.  We obtained upper bounds for the energy scale $\Lambda$ associated to those effective operators, $\Lambda \leq  4.7 v_{\chi}$ required in order to explain all light masses simultaneously. \\

After that, we proceeded to study phenomenological consequences of the model in the flavour sector, specifically we started by calculating all the leading order contributions to the anomalous magnetic moment of the muon $g-2$ by considering the interactions of heavy scalars, both neutral and charged, heavy fermions as the charged lepton $E$ and TeV Majorana neutrinos, all of them resulting in negative contributions. However, the interaction between the SM $W^{+}$ gauge boson with those exotic neutrinos provided the only positive contribution capable to explain the experimental value obtained by Fermilab as long as the Yukawa couplings of the neutral leptons acquire values larger than $4.5$. \\

Finally, we investigated the leading order effects generated by the model regarding the $B$ meson anomalies. We considered the interaction of the $Z_{1}^{\mu}$ gauge boson to be equal to the SM $Z^{\mu}$, while checking that the interactions with the $Z_{2}^{\mu}$ boson generated new physics contributions encapsulated in effective Wilson coefficients. We found that from the recent LHCb measurements of $R_{K^{*}}$ at intermediate $q^2$, the model can explain the anomalies at the $2\sigma$ level alone, favouring in particular $\tan\beta\approx15$ from neutrino trident production constraints, which can accommodate the fermion mass hierarchies within the model. Furthermore, from $B_{s}-\overline{B}_{s}$ mixing, we found a strong upper bound on the $\upsilon_{\chi}$ VEV of $5.7$ TeV.

\begin{acknowledgements}
The work of CS is supported by the Excellent Postdoctoral Program  of Jiangsu Province grant No. 2023ZB891.
\end{acknowledgements}

\appendix
\label{appendix}
\section{Scalar boson masses}\label{scalars}
First, the most general scalar potential allowed by the symmetries of the model is given by \cite{orig},
\begin{align}
V &= \mu_{1}^{2}\phi_{1}^{\dagger}\phi_{1} + \mu_{2}^{2}\phi_{2}^{\dagger}\phi_{2} + \mu_{\chi}^{2}\chi^{*}\chi + \frac{f}{\sqrt{2}}\left(\phi_{1}^{\dagger}\phi_{2}\chi ^{*} + \mathrm{\text{H.C.}} \right)  \nonumber\\ &
 + \lambda_{1}\left(\phi_{1}^{\dagger}\phi_{1}\right)^{2} + \lambda_{2}\left(\phi_{2}^{\dagger}\phi_{2}\right)^{2} 
 + \lambda_{3}\left(\chi^{*}\chi \right)^{2}  \nonumber\\	&
 + \lambda_{5}\left(\phi_{1}^{\dagger}\phi_{1}\right) \left(\phi_{2}^{\dagger}\phi_{2}\right)
 + \lambda'_{5}\left(\phi_{1}^{\dagger}\phi_{2}\right)\left(\phi_{2}^{\dagger}\phi_{1}\right)	\nonumber\\ 
 & + \lambda_{6} \left(\phi_{1}^{\dagger}\phi_{1}\right) \left(\chi^{*}\chi \right)   + \lambda_{7}\left(\phi_{2}^{\dagger}\phi_{2}\right) \left(\chi^{*}\chi \right).
\end{align}
\noindent
This potential generates a mass matrix for charged, CP-even and CP-odd scalars after SSB takes place. For charged scalars, the mass matrix written in the basis $(\phi^{\pm}_{1},\phi^{\pm}_{2})$ is,

\begin{equation}
\mathit{M}_{\mathrm{C}}^{2} = \frac{1}{4}
\begin{pmatrix}
	-f\dfrac{v_{\chi}v_{2}}{v_{1}}-\lambda_{5}'{{v_{2}}^{2}} & 
	 fv_{\chi}+\lambda_{5}'v_{1}v_{2}		\\
	 fv_{\chi}+\lambda_{5}'v_{1}v_{2}		&
	-f\dfrac{v_{\chi}v_{1}}{v_{2}}-\lambda_{5}'{{v_{1}}^{2}}
\end{pmatrix},
\end{equation}
\noindent
with the respective rotation matrix connecting interaction states to mass eigenstates $\boldsymbol{H}^{\pm}=(G_{W}^{\pm},H^{\pm})$ given by,
\begin{equation}
\begin{split}
\boldsymbol{\phi}^{\pm} &= \mathbb{R}_{\phi}\boldsymbol{H}^{\pm}, \\
\begin{pmatrix}
\phi_{1}^{\pm}	\\
\phi_{2}^{\pm}
\end{pmatrix} &= 
\begin{pmatrix}
c_{\beta}	&	s_{\beta}	\\
-s_{\beta}	&	c_{\beta}
\end{pmatrix}
\begin{pmatrix}
\boldsymbol{H}^{\pm}	\\
G_{W}^{\pm}
\end{pmatrix},
\end{split}
\end{equation}
\noindent
where $s_{\beta}=\sin\beta$, $c_{\beta}=\cos\beta$, $t_{\beta}=s_{\beta}/c_{\beta}=v_{1}/v_{2}$ with $v_{1}>v_{2}$ and the corresponding mass eigenvalues are,
\begin{equation}
\begin{split}
m_{G_{W}^{\pm}}^{2} &= 0,	\\
m_{\boldsymbol{H}^{\pm}}^{2} &= -\frac{1}{4}\frac{f v_{\chi}}{s_{\beta}c_{\beta}} -\frac{1}{4}\lambda_{5}' v^2,
\end{split}
\end{equation}
\noindent
and $G_{W}^{\pm}$ is identified as the would-be Goldstone boson eaten by the the $W$ gauge boson. \\

Regarding the neutral bosons, the CP-odd scalar bosons of the model  $\boldsymbol{\eta}=(\eta_{1},\eta_{2},\zeta_{\chi})$ mix together according to the mass matrix,

\begin{equation}
\mathit{M}_{\mathrm{I}}^{2} = -\frac{f}{4}
\begin{pmatrix}
\dfrac{ {v_{2}}\, {v_{\chi}}}{ {v_{1}}} & - {v_{\chi}} &  {v_{2}}\\
- {v_{\chi}} & \dfrac{ {v_{1}}\, {v_{\chi}}}{ {v_{2}}} & - {v_{1}}\\
 {v_{2}} & - {v_{1}} & \dfrac{ {v_{1}}\, {v_{2}}}{ {v_{\chi}}}
\end{pmatrix},
\end{equation}
\noindent
with mass eigenstates $\boldsymbol{A}=(A^{0},G_{Z},G'_{Z})$ containing only one physical pseudoscalar particle identified as $A^{0}$, with mass given by,

\begin{equation}
\begin{split}
m_{A^{0}}^{2}	&=	-\frac{1}{4}\frac{f v_{\chi}}{s_{\beta}c_{\beta}c_{\gamma}^{2}} \approx -\frac{1}{4}\frac{f v_{\chi}}{s_{\beta}c_{\beta}},
\end{split}
\end{equation}
\noindent
where $t_{\gamma}=\tan\gamma=v s_{\beta}c_{\beta}/v_{\chi} \ll 1$ and, $G_{Z}$, $G'_{Z}$ correspond to the massless Goldstone bosons eaten by the $Z$ and $Z'$ physical gauge bosons, respectively. The CP-even bosons mix together according to,
\begin{equation}
\begin{split}
\boldsymbol{\eta} &= \mathbb{R}_{\eta}\boldsymbol{A}, \\
\begin{pmatrix}
\eta_{1}	\\
\eta_{2}	\\
\zeta_{\chi}
\end{pmatrix} &= 
\begin{pmatrix}
c_{\beta}	&	s_{\beta}	&	0	\\
-s_{\beta}	&	c_{\beta}	&	0	\\
0	&	0	&	1
\end{pmatrix}\!\!\! 
\begin{pmatrix}
c_{\gamma}	&	0	&	-s_{\gamma}	\\
0	&	1	&	0	\\
s_{\gamma}	&	0	&	c_{\gamma}
\end{pmatrix}\!\!\! 
\begin{pmatrix}
A^{0}	\\
G_{Z}	\\
G'_{Z}
\end{pmatrix} \\ &= 
\begin{pmatrix}
c_{\beta}c_{\gamma}	&	s_{\beta}	&	-c_{\beta}s_{\gamma}	\\
-s_{\beta}c_{\gamma}	&	c_{\beta}	&	s_{\beta}s_{\gamma}	\\
s_{\gamma}	&	0	&	c_{\gamma}
\end{pmatrix}\!\!\! 
\begin{pmatrix}
A^{0}	\\
G_{Z}	\\
G'_{Z}
\end{pmatrix}.
\end{split}
\end{equation}
\noindent

Lastly, the CP-even scalar bosons of the model $\boldsymbol{h}=(h_{1},h_{2},\xi_{\chi})$ give rise to the following mass matrix,
\footnotesize
\begin{align}
\mathit{M}_{\mathrm{R}}^{2} &= \begin{pmatrix}
	\lambda_{1} {v_{1}^{2}}-\dfrac{1}{4}\dfrac{f v_{\chi} v_{2}}{v_{1}} &
	\hat{\lambda}_{5} { v_{1} v_{2}}+\dfrac{1}{4}{f v_{\chi}} &
	\dfrac{1}{4}\lambda_{6}{ v_{1} v_{\chi}}+\dfrac{1}{4}{f v_{2}}		\\
	\hat{\lambda}_{5} { v_{1} v_{2}}+\dfrac{1}{4}{f v_{\chi}} &
	\lambda_{2} {v_{2}^{2}}-\dfrac{1}{4}\dfrac{f v_{\chi} v_{1}}{v_{2}} & 
	\dfrac{1}{4}\lambda_{7}{ v_{2} v_{\chi}}+\dfrac{1}{4}{f v_{1}}		\\
	\dfrac{1}{4}\lambda_{6}{ v_{1} v_{\chi}}+\dfrac{1}{4}{f v_{2}}	& 
	\dfrac{1}{4}\lambda_{7}{ v_{2} v_{\chi}}+\dfrac{1}{4}{f v_{1}} & 
	\lambda_{3} {v_{\chi}^{2}}-\dfrac{1}{4}\dfrac{f v_{1} v_{2}}{v_{\chi}}
\end{pmatrix}, \nonumber
\end{align}
\normalsize
which after diagonalization to the mass basis $\boldsymbol{H}=(H,h,H_{\chi})$ has the following eigenvalues,
\begin{align}
m_{h}^{2} &\approx \left( \tilde{\lambda}_{1}c_{\beta}^{4}+2\tilde{\lambda}_{5}c_{\beta}^{2}s_{\beta}^{2}+\tilde{\lambda}_{2}s_{\beta}^{4} \right){v^{2}},	\\
m_{H}^{2} &\approx -\frac{f v_{\chi}}{4s_{\beta}c_{\beta}},	\\
m_{H_{\chi}}^{2} &\approx \lambda_{3} {v_{\chi}^{2}}.
\end{align}
where the tilded constants defined as,
 \begin{align}
\tilde{\lambda}_{1} &= \lambda_{1}-\frac{\lambda_{6}^2}{4 \lambda_{3}}-\frac{\lambda_{7}^2}{4 \lambda_{3} t_{\beta}^2},\\
\tilde{\lambda}_{2} &= \lambda_{2}-\frac{\lambda_{6}^2 t_{\beta}^2}{4 \lambda_{3}}-\frac{\lambda_{7}^2}{4 \lambda_{3}},\\
\tilde{\lambda}_{5} &= \hat{\lambda}_{5}-\frac{\lambda_{6}^2 t_{\beta}}{2 \lambda_{3}}-\frac{\lambda_{7}^2}{2 \lambda_{3} t_{\beta}},
\end{align}
\noindent
and the rotation matrix written as the product of three matrices is given by,
\begin{equation}
\begin{split}
\begin{pmatrix}
h_{1}	\\
h_{2}	\\
\xi_{\chi}
\end{pmatrix} \!\! &= \!\!
\begin{pmatrix}
1	&	0	&	0	\\
0	&	c_{23}	&	s_{23}	\\
0	&	-s_{23}	&	c_{23}
\end{pmatrix}  
\begin{pmatrix}
c_{13}	&	0	&	s_{13}	\\
0	&	1	&	0	\\
-s_{13}	&	0	&	c_{13}
\end{pmatrix}\!\!\!\! 
\begin{pmatrix}
c_{\alpha}	&	s_{\alpha}	&	0	\\
-s_{\alpha}	&	c_{\alpha}	&	0	\\
0	&	0	&	1
\end{pmatrix}
\begin{pmatrix}
H	\\
h	\\
H_{\chi}
\end{pmatrix},
\end{split}
\end{equation}
where the mixing angles are defined as,
\begin{align}
s_{23} &= \frac{\lambda_{7}c_{\beta}v}{2\lambda_{3}v_{\chi}}, & 
s_{13}  &= \frac{\lambda_{6}s_{\beta}v}{2\lambda_{3}v_{\chi}}, &
t_{2\alpha} &= \frac{f v_{\chi} + 2\tilde{\lambda}_{5}s_{\beta}c_{\beta}v^{2}}{f v_{\chi} + 2t_{2\beta}(\tilde{\lambda}_{1}s_{\beta}^{2}-\tilde{\lambda}_{2}c_{\beta}^{2})v^{2}} t_{2\beta}.
\end{align}
As it is shown in section \ref{quarks}, the top quark mass is proportional to $v_{1}$ while the down quark is proportional to $v_{2}$, so their mass difference can be understood by the value of each VEV. Thus, it is appropriate to consider some approximations that can be done assuming $s_{\beta}\approx 1$,
\begin{align}
c_{\beta}&\approx 0,& 
t_{\alpha}&\approx t_{\beta},&
s_{13}&\approx \dfrac{\lambda_{6}vs_{\beta}}{2\lambda_{3}v_{\chi}},& 
s_{23}&\approx 0,
\end{align}
which reduces the mixing matrix for the CP-even states as,
\begin{equation}
\begin{split}
\boldsymbol{h} &= \mathbb{R}_{h}\boldsymbol{H}, \\
\begin{pmatrix}
h_{1}	\\
h_{2}	\\
\xi_{\chi}
\end{pmatrix} \!\! &= \!\!
\begin{pmatrix}
c_{13}	&	0	&	s_{13}	\\
0	&	1	&	0	\\
-s_{13}	&	0	&	c_{13}
\end{pmatrix}\!\!\! 
\begin{pmatrix}
c_{\alpha}	&	s_{\alpha}	&	0	\\
-s_{\alpha}	&	c_{\alpha}	&	0	\\
0	&	0	&	1
\end{pmatrix}\!\!\! 
\begin{pmatrix}
H	\\
h	\\
H_{\chi}
\end{pmatrix} \\ &= 
\begin{pmatrix}
c_{\beta}c_{13}	&	s_{\beta}c_{13}	& s_{13}	\\
-s_{\beta}  	&	c_{\beta}	    & 0	\\
-c_{\beta}s_{13}&	s_{\beta}s_{13}& c_{13}
\end{pmatrix}\!\!\! 
\begin{pmatrix}
H	\\
h	\\
H_{\chi}
\end{pmatrix}.
\end{split}
\end{equation}

From the three CP-even physical states, the lightest one, $h$, is identified as the SM Higgs boson, while $H$ and $H_{\chi}$ are heavier and yet unobserved particles whose mass depends on the $U(1)_{X}$ symmetry breaking scale $v_{\chi}$ similarly to $A^{0}$ and $\boldsymbol{H}^{\pm}$. Therefore, heavy scalars have approximately the same mass, $m_{H}\approx m_{H_{\chi}}\approx m_{A^{0}}\approx m_{\boldsymbol{H}^{\pm}}$ and according to the lower bound on charged scalar given by \cite{chargedbound} we can assume a lower bound for their masses around $800$ GeV.

\section{Gauge boson masses}\label{gaugebosons}
After SSB, the charged gauge bosons $W_{\mu }^{\pm}=(W_{\mu }^{1}\mp W_{\mu }^{2})/\sqrt{2}$ acquire a mass given by $m_{W}=g\,v/2$. Regarding the neutral gauge bosons of the model, they are arranged in the basis $(W_{\mu}^{3},B_{\mu},Z'_{\mu})$, producing the following mass matrix,
\begin{equation*}
M_0^2=\frac{1}{4}\begin{pmatrix}
g^2 v^2 &-gg' v^2   &-\frac{2}{3}gg_Xv^2(1+c_{\beta} ^2)   \\ 
&&\\
* &  g'^2v^2 &  \frac{2}{3}g'g_X v^2(1+c_{\beta}^2)   \\
 &&\\
* &*  &  \frac{4}{9}g_X^2 v_{\chi}^2\left[1+(1+3 c_{\beta}^2)\frac{v^{2}}{v_{\chi}^{2}}\right] \\
\end{pmatrix},
\end{equation*} 
\noindent
and their states mix together to form mass eigenstates $(A_{\mu},Z_{\mu}^{1},Z_{\mu}^{2})$,
\begin{equation}
\begin{pmatrix}
W_{\mu}^{3}\\
B_{\mu}\\
Z'_{\mu}
\end{pmatrix}
=
\begin{pmatrix}
s_{W}	&	c_{W}	&	0	\\
c_{W}	&	-s_{W}	&	0	\\
0	&	0	&	1
\end{pmatrix}\!\!\! 
\begin{pmatrix}
1	&	0	&	0	\\
0	&	c_{Z}	&	-s_{Z}	\\
0	&	s_{Z}	&	 c_{Z}
\end{pmatrix}\!\!\! 
\begin{pmatrix}
A_{\mu}\\
Z_{\mu}^{1}\\
Z_{\mu}^{2}
\end{pmatrix},
\end{equation}
with the Weinberg angle defined as $t_{W}=s_{W}/c_{W}=\tan\theta_{W}=g'/g$ and $\sin\theta_{Z}=s_{Z}$,
\begin{equation}
s_{Z} \approx (1+s_{\beta}^{2})
\frac{2g_{X}c_{W}}{3g}\frac{M_{Z}}{M_{Z'}} \approx 
\frac{2v}{v_{\chi}}\lesssim 10^{-2},
\end{equation}
and where in the last approximation we have assumed $t_{\beta} \gg 1$ and $\theta_{Z}$ as a small mixing angle between $Z$ and $Z'$ gauge boson as indicated by LEP data \cite{ALEPH:2005ab}. Then, the masses for the neutral gauge bosons are given by,
\begin{equation}
M_{1}\approx M_{Z } = \frac{gv}{2c_{W}}, \qquad
M_{2} \approx M_{Z'} \approx \frac{g_{X}v_{\chi}}{3}, \label{mzp}
\end{equation}

and the total mixing of the neutral gauge bosons can be written as,
\begin{equation}
\begin{pmatrix}
W_{\mu}^{3}\\
B_{\mu}\\
Z'_{\mu}
\end{pmatrix}
\approx
\begin{pmatrix}
s_{W}	&	c_{W}c_{Z}	&	-c_{W}s_{Z}	\\
c_{W}	&	-s_{W}c_{Z}	&	s_{W}s_{Z}	\\
0	&	s_{Z}	&	c_{Z}
\end{pmatrix}\!\!\! 
\begin{pmatrix}
A_{\mu}\\
Z_{\mu}^{1}\\
Z_{\mu}^{2}
\end{pmatrix}.
\end{equation}

\section{Charged leptons matrix rotation}\label{chargedlpetons}

The rotations associated to the charged leptons used in section \ref{leptons} are defined by,

\begin{align}
\mathbb{V}^{E}_{L\; 1} & \approx
\begin{pmatrix}
1	&	0	&	0 & \frac{v_{1}q_{11}}{\sqrt{2}m_{E}} \\
 0&	1 & 0 &  \frac{v_{1}q_{12}}{\sqrt{2}m_{E}} \\
0&0  & 1 & r_{3} \\
-\frac{v_{1}q_{11}}{\sqrt{2}m_{E}}&	-\frac{v_{1}q_{12}}{\sqrt{2}m_{E}}	& -r_{3} &	1
\end{pmatrix}, &
\mathbb{V}^{E}_{L\; 2} & \approx
\begin{pmatrix}
c_{e\mu}	&	s_{e\mu}	& r_{1} & 0 \\
-s_{e\mu}	&	c_{e\mu}	& r_{2} & 0 \\
-r_{1}c_{e\mu}+r_{2}s_{e\mu} & -r_{2}c_{e\mu} -r_{1}s_{e\mu} & 1 & 0 \\
0&	0	& 0 &	1
\end{pmatrix}, \nonumber\\
\mathbb{V}^{E}_{R\; 1} & \approx\begin{pmatrix}
1 & 0 & 0 & \frac{\Omega^{\ell}_{41}v_{1}v_{2}}{2m_{E}\Lambda^{2}} \\
0 & 1 & 0 & t_{1} \\
0 & 0 & 1 &   \frac{\Omega^{\ell}_{43}v_{1}v_{2}}{2m_{E}\Lambda^{2}}\\
-\frac{\Omega^{\ell}_{41}v_{1}v_{2}}{2m_{E}\Lambda^{2}} & -t_{1} &  -\frac{\Omega^{\ell}_{43}v_{1}v_{2}}{2m_{E}\Lambda^{2}} & 1 \\
\end{pmatrix},&
\mathbb{V}^{E}_{R\; 2} & \approx\begin{pmatrix}
c_{e\tau} & -c_{e\tau}t_{2} & s_{e\tau} & 0 \\
t_{2} & 1 & 0 & 0 \\
-s_{e\tau} & -s_{e\tau}t_{2} & c_{e\tau} & 0 \\
0 & 0 & 0 & 1
\end{pmatrix} \label{VRE2},
\end{align}
where we have introduced the following definitions in order to simplify the rotation matrices,
\begin{align}
    r_{1}=&\frac{(s_{e\tau}\Omega^{\ell}_{11}+c_{e\tau}\Omega^{\ell}_{13})v_{2}v_{\chi}^{3}+\sqrt{2}m_{\mu}v_{\chi}^{3}\Omega^{\ell}_{32}s_{e\mu}}{4\Lambda^{3}m_{\tau}}, &
    r_{2}=&\frac{(s_{e\tau}\Omega^{\ell}_{21}+c_{e\tau}\Omega^{\ell}_{23})v_{2}v_{\chi}^{3}+\sqrt{2}m_{\mu}v_{\chi}^{3}\Omega^{\ell}_{32}s_{e\mu}}{4\Lambda^{3}m_{\tau}}, \nonumber\\
    r_{3}=&\frac{v_{1}v_{2}}{2\Lambda m_{E}}\left( \frac{\Omega^{\ell}_{34}v_{\chi}^{3}}{2\Lambda^{2}v_{2}}+\frac{m_{\tau}(\Omega^{\ell}_{41}s_{e\tau}-\Omega^{\ell}_{43}c_{e\tau})}{m_{E}}\right), &
    t_{1}=&\frac{\Omega^{\ell}_{42}v_{1}v_{2}v_{\chi}}{2m_{E}\Lambda^{2}}+\frac{v_{1}m_{\mu}}{\sqrt{2}m_{E}^{2}}(q_{11}s_{e\mu}+q_{12}c_{e\mu}) ,\nonumber\\
    t_{2}=& \frac{v_{2}v_{\chi}^{3}}{4\Lambda^{3}m_{\mu}}(s_{e\mu}(\Omega^{\ell}_{13}s_{e\tau}-\Omega^{\ell}_{11}c_{e\tau})  +c_{e\mu}(\Omega^{\ell}_{21}c_{e\tau}+\Omega^{\ell}_{23}s_{e\tau})), \label{r1r2r3t1t2}
\end{align}
\noindent
 with $t_{e\mu}=\eta/h$ and $t_{e\tau}=\zeta/H$.
\section{Up-like quarks parameters}\label{upquarks}
The rotations for the up-like quarks are defined as,

\begin{align}
\mathbb{V}^{U}_{L\; 1} & \approx
\begin{pmatrix}
1	&	0	&	0 & \frac{v_{2}r_{1}^{-}}{\sqrt{2}m_{\mathcal{T}}} \\
 0&	1 & 0 &  \frac{v_{1}r_{2}^{-}}{\sqrt{2}m_{\mathcal{T}}} \\
0&0  & 1 & \frac{v_{2}v_{\chi}r_{3}^{-}}{2m_{\mathcal{T}}\Lambda} \\
-\frac{v_{2}r_{1}^{-}}{\sqrt{2}m_{\mathcal{T}}}&	- \frac{v_{1}r_{2}^{-}}{\sqrt{2}m_{\mathcal{T}}}	& -\frac{v_{2}v_{\chi}r_{3}^{-}}{2m_{\mathcal{T}}\Lambda} &	1
\end{pmatrix}, &
\mathbb{V}^{U}_{L\; 2} & \approx
\begin{pmatrix}
c_{uc}	&	s_{uc}	& r_{1}^{U} & 0 \\
-s_{uc}	&	c_{uc}	& r_{2}^{U} & 0 \\
-r_{1}^{U}c_{uc}+r_{2}^{U}s_{uc} & -r_{2}^{U}c_{uc} -r_{1}^{U}s_{uc} & 1 & 0 \\
0&	0	& 0 &	1
\end{pmatrix},  \nonumber\\
\mathbb{V}^{U}_{R} & \approx\begin{pmatrix}
c_{ut} & 0 & s_{ut} & 0 \\
0 & -s_{\alpha} & 0 & c_{\alpha} \\
-s_{ut} & 0 & c_{ut} & 0 \\
0 & c_{\alpha} & 0 & s_{\alpha}
\end{pmatrix}, \label{VRU}
\end{align}
\noindent
with the $r_{1}^{U}$ and $r_{2}^{U}$ parameters defined as,
\begin{align}
    r_{1}^{U}&=\frac{v_{\chi}v_{2}^{2}r_{3}^{+}r_{1}^{+}}{2\sqrt{2}m_{t}^{2}\Lambda} + \frac{v_{\chi}v_{1}(\Omega_{11}^{U}s_{ut}+\Omega_{13}^{U}c_{ut})}{2 m_{t}\Lambda}, &
    r_{2}^{U}&=\frac{v_{1}v_{2}v_{\chi}(\Omega_{21}^{U}s_{ut}+\Omega_{23}^{U}c_{ut} + r_{3}^{+}r_{2}^{+})}{2\sqrt{2}m_{t}^{2}\Lambda},
\end{align}

where we have introduced the following set of rotated parameters,
\begin{align}
\begin{pmatrix}
r_{1}^{+} \\
r_{1}^{-}
\end{pmatrix}
&=\begin{pmatrix}
\cos\alpha & -\sin\alpha \\
\sin\alpha & \cos\alpha
\end{pmatrix}
\begin{pmatrix}
(h_{2}^{T})_{1} \\
(h_{2}^{U})_{12}
\end{pmatrix}, &
\begin{pmatrix}
r_{2}^{+} \\
r_{2}^{-}
\end{pmatrix}
&=\begin{pmatrix}
\cos\alpha & -\sin\alpha \\
\sin\alpha & \cos\alpha
\end{pmatrix}
\begin{pmatrix}
(h_{1}^{T})_{2} \\
(h_{1}^{U})_{22}
\end{pmatrix}, \nonumber\\
\begin{pmatrix} 
r_{3}^{+} \\
r_{3}^{-}
\end{pmatrix}
&=\begin{pmatrix}
\cos\alpha & -\sin\alpha \\
\sin\alpha & \cos\alpha
\end{pmatrix}
\begin{pmatrix}
\Omega_{34}^{U} \\
\Omega_{32}^{U}
\end{pmatrix},&
    \tan\alpha&=\frac{h_{\chi}^{T}}{(h_{\chi}^{U})_{2}}.
\end{align}
\\
and the $\theta_{uc(t)}$ angles are defined as,
\begin{align}
    \tan\theta_{uc}&=t_{uc}=\frac{v_{2}r_{1}^{+}}{v_{1}r_{2}^{-}}=\frac{v_{2}}{v_{1}}\frac{(h_{2}^{T})_{1}(h_{\chi}^{U})_{2}-(h_{2}^{U})_{12}h_{\chi}^{T}}{(h_{1}^{T})_{2}(h_{\chi}^{U})_{2}-(h_{1}^{U})_{22}h_{\chi}^{T}}, &   \tan\theta_{ut}&=\frac{(h_{1}^{U})_{13}}{(h_{1}^{U})_{33}}. \label{thetauc}
\end{align}

\section{Down-like quarks parameters}\label{downquarks}

The parameters appearing in Eqs.(\ref{mdown}-\ref{mJ2}) are defined as,
\small
\begin{align}
    \xi_{11}&=\Omega^{D}_{11\; 2}+\Omega^{D\; 2}_{12}+\Omega^{D\; 2}_{13}-\frac{\left(\Omega^{D}_{11} (h_{2}^{D})_{31}+\Omega^{D}_{12} (h_{2}^{D})_{32}+\Omega^{D}_{13} (h_{2}^{D})_{33}\right){}^2}{\left(Y_d\right)_{3,1}^2+\left(Y_d\right)_{3,2}^2+\left(Y_d\right)_{3,3}^2}, \nonumber \\ 
    \xi_{22}&=\Omega^{D\; 2}_{21}+\Omega^{D\; 2}_{22}+\Omega^{D\; 2}_{23} -\frac{\left(\Omega^{D}_{21} (h_{2}^{D})_{31}+\Omega^{D}_{22} (h_{2}^{D})_{32}+\Omega^{D}_{23} (h_{2}^{D})_{33}\right){}^2}{\left(Y_d\right)_{3,1}^2+\left(Y_d\right)_{3,2}^2+\left(Y_d\right)_{3,3}^2}, \nonumber\\
    \xi_{12}&=\Omega^{D}_{11} \Omega^{D}_{21}+\Omega^{D}_{12} \Omega^{D}_{22}+\Omega^{D}_{13} \Omega^{D}_{23} -\frac{\left[\left(\Omega^{D}_{11} (h_{2}^{D})_{31}+\Omega^{D}_{12} (h_{2}^{D})_{32}+\Omega^{D}_{13} (h_{2}^{D})_{33}\right) \left(\Omega^{D}_{21} (h_{2}^{D})_{31}+\Omega^{D}_{22} (h_{2}^{D})_{32}+\Omega^{D}_{23} (h_{2}^{D})_{33}\right) \right]}{  (\left(Y_d\right)_{3,1}^2+\left(Y_d\right)_{3,2}^2+\left(Y_d\right)_{3,3}^2)},\nonumber \\
    \rho&=(h_{\chi}^{J})_{11})^{2}+((h_{\chi}^{J})_{12})^{2}+((h_{\chi}^{J})_{21})^{2}+((h_{\chi}^{J})_{25}, \nonumber\\
    \eta&=(h_{\chi}^{J})_{12} (h_{\chi}^{J})_{21}-(h_{\chi}^{J})_{11} (h_{\chi}^{J})_{12}.
\end{align}
\normalsize

Furthermore, the rotation matrix for left-handed down-like quarks can be written as $ \mathbb{V}^{D}_{L} \approx \mathbb{V}^{D}_{L\; 1} \mathbb{V}^{D}_{L\; 2}$, where each matrix reads,

\small
\begin{align}
    \mathbb{V}^{D}_{L\; 1} & \approx \begin{pmatrix}
    1 & 0 & 0 & \frac{v_{1}v_{\chi}}{2} \kappa^{L}_{12} & \frac{v_{1}v_{\chi}}{2} \kappa^{L}_{11}\\
    0 & 1 & 0 & \frac{v_{2}v_{\chi}}{2}\kappa^{L}_{22} & \frac{v_{2}v_{\chi}}{2}\kappa^{L}_{21}\\
    0 & 0 & 1 & \frac{v_{2}v_{\chi}^{2}}{2\sqrt{2}\Lambda }\kappa^{L}_{32} & \frac{v_{2}v_{\chi}^{2}}{2\sqrt{2}\Lambda }\kappa^{L}_{31} \\
   -\frac{v_{1}v_{\chi}}{2} \kappa^{L}_{12} & -\frac{v_{2}v_{\chi}}{2}\kappa^{L}_{22} & -\frac{v_{2}v_{\chi}^{2}}{2\sqrt{2}\Lambda }\kappa^{L}_{32} & 1 & 0 \\
   -\frac{v_{1}v_{\chi}}{2} \kappa^{L}_{11} & -\frac{v_{2}v_{\chi}}{2}\kappa^{L}_{21} & -\frac{v_{2}v_{\chi}^{2}}{2\sqrt{2}\Lambda }\kappa^{L}_{31}& 0 & 1
    \end{pmatrix},\label{VLD1} \\
    \mathbb{V}^{D}_{L\; 2} & \approx \begin{pmatrix}
c_{ds}	&	s_{ds}	& r_{1}^{D} & 0 & 0 \\
-s_{ds}	&	c_{ds}	& r_{2}^{D} & 0 & 0\\
-r_{1}^{D}c_{ds}+r_{2}^{D}s_{ds} & -r_{2}^{D}c_{ds} -r_{1}^{D}s_{ds} & 1 & 0 & 0\\
0&	0	& 0 &	1& 0 \\
0 & 0 & 0 & 0 & 1
\end{pmatrix}, \label{VLD2}
\end{align}
\normalsize
\noindent
with the parameters
\small
\begin{align}
    &\kappa^{L}_{ij}=\frac{1}{m_{\mathcal{J}^{1}}m_{\mathcal{J}^{2}}}(-Y_{i5}(h_{\chi}^{J})_{j1}+ Y_{i4}(h_{\chi}^{J})_{j2}), \\
    &r_{1}^{D}=\frac{v_{\chi}v_{2}^{2}}{2\sqrt{2}\Lambda m_{b}^{2}}(\Omega_{11}^{D}(h_{2}^{D})_{31}+\Omega_{12}^{D}(h_{2}^{D})_{32}+\Omega_{13}^{D}(h_{2}^{D})_{33}), \label{r1D} \\
    &r_{2}^{D}=\frac{v_{\chi}v_{1}v_{2}}{2\sqrt{2}\Lambda m_{b}^{2}}(\Omega_{21}^{D}(h_{2}^{D})_{31}+\Omega_{22}^{D}(h_{2}^{D})_{32}+\Omega_{23}^{D}(h_{2}^{D})_{33}), \\
    &\tan \theta_{ds}=t_{ds}=\frac{v_{1}v_{2}^{3}\xi_{12}}{v_{1}^{2}v_{2}^{2}\xi_{22}-4m_{b}^{2}m_{d}^{2}}, \label{thetads}
\end{align}
\normalsize
\noindent
and $Y_{ij}$ represents the Yukawa couplings in the $(\mathbb{M}_{D})_{ij}$ entry of the mass matrix. For instance, $Y_{15}=(h_{1}^{J})_{2}$ or $Y_{34}=\Omega_{34}^{D}$. We can also get the rotation matrix elements through the Cabibbo-Kobayashi-Maskawa (CKM) matrix, which is defined by $\mathbb{V}=\mathbb{V}_{L}^{U\dagger}\mathbb{V}_{L}^{D}$, in such a way that the down-like quarks phenomenology can be related to the $\theta_{uc}$ angle in Eq. (\ref{thetauc}) as well. \\

On the other hand, the rotation matrix for right-handed down-like quarks has a more complicated structure. It can also be written as $ \mathbb{V}^{D}_{R}\approx \mathbb{V}^{D}_{R\; 1} \mathbb{V}^{D}_{R\; 2}$
 where the rotation matrices are defined as,
 
 \begin{align}
    \mathbb{V}^{D}_{R\; 1}  &\approx \begin{pmatrix}
    1 & 0 & 0 & -\frac{v_{2}v_{\chi}^{2}}{2\sqrt{2}\Lambda}\kappa^{R}_{11} & -\frac{v_{2}v_{\chi}^{2}}{2\sqrt{2}\Lambda}\kappa^{R}_{12}\\
    0 & 1 & 0 & -\frac{v_{2}v_{\chi}^{2}}{2\sqrt{2}\Lambda}\kappa^{R}_{21} & -\frac{v_{2}v_{\chi}^{2}}{2\sqrt{2}\Lambda}\kappa^{R}_{22}\\
    0 & 0 & 1 & -\frac{v_{2}v_{\chi}^{2}}{2\sqrt{2}\Lambda}\kappa^{R}_{31} & -\frac{v_{2}v_{\chi}^{2}}{2\sqrt{2}\Lambda}\kappa^{R}_{22} \\
    \frac{v_{2}v_{\chi}^{2}}{2\sqrt{2}\Lambda}\kappa^{R}_{11} & \frac{v_{2}v_{\chi}^{2}}{2\sqrt{2}\Lambda}\kappa^{R}_{21} & \frac{v_{2}v_{\chi}^{2}}{2\sqrt{2}\Lambda}\kappa^{R}_{31} & 1 & 0 \\
   \frac{v_{2}v_{\chi}^{2}}{2\sqrt{2}\Lambda}\kappa^{R}_{12} & \frac{v_{2}v_{\chi}^{2}}{2\sqrt{2}\Lambda}\kappa^{R}_{22} & \frac{v_{2}v_{\chi}^{2}}{2\sqrt{2}\Lambda}\kappa^{R}_{32} & 0 & 1
    \end{pmatrix},\label{VRD1}\\
    &\mathbb{V}^{D}_{R\; 2}  \approx \begin{pmatrix}
    c_{13} & 0 & -s_{13} & 0 & 0\\
    0 & 1 & 0 & 0 & 0\\
    s_{13} & 0 & c_{13} & 0 & 0 \\
    0 & 0 & 0 & 1 & 0 \\
    0 & 0 & 0 & 0 & 1
    \end{pmatrix} 
\begin{pmatrix}
1 & 0 & 0 & 0& 0\\
0 & c_{23} & -s_{23} & 0& 0\\
0 & s_{23} & c_{23} & 0 & 0\\
 0 & 0 & 0 & 1& 0 \\
  0 & 0 & 0 & 0 & 1
\end{pmatrix} 
\begin{pmatrix}
c_{12} & s_{12} & 0 & 0 & 0\\
-s_{12} & c_{12} & 0 & 0& 0\\
0 &0 & 1  & 0 & 0\\
 0 & 0 & 0 & 1& 0 \\
  0 & 0 & 0 & 0 & 1
\end{pmatrix}, \label{VRD2}
\end{align}
\normalsize
where 
\begin{align}\kappa^{R}_{ij}&=\Omega_{4i}^{D}(h_{\chi}^{J})_{1j} +
\Omega_{5i}^{D}(h_{\chi}^{J})_{2j}, \\    t_{13}&=\frac{(h_{2}^{D})_{31}}{(h_{2}^{D})_{33}}, \\
    t_{23}&=\frac{(h_{2}^{D})_{32}}{\sqrt{((h_{2}^{D})_{31})^{2}+((h_{2}^{D})_{32})^{2}+((h_{2}^{D})_{33})^{2}}}, \\
    t_{12}&\approx s_{12}\nonumber\\
    &=\frac{v_{\chi}^{2}}{4\Lambda^{2}m_{s}^{2}}\Bigg( \frac{\Omega^{D}_{22} v_1^2 \left(\Omega^{D}_{21} (h_{2}^{D})_{33}-\Omega^{D}_{23} (h_{2}^{D})_{31}\right)}{\sqrt{\left(Y_d\right)_{3,1}^2+\left(Y_d\right)_{3,2}^2+\left(Y_d\right)_{3,3}^2}}+\frac{\Omega^{D}_{12} v_2^2 \left(\Omega^{D}_{11} (h_{2}^{D})_{33}-\Omega^{D}_{13} (h_{2}^{D})_{31}\right)}{\sqrt{\left(Y_d\right)_{3,1}^2+\left(Y_d\right)_{3,2}^2+\left(Y_d\right)_{3,3}^2}}\nonumber \\
    &+s_{23} [\left(\left(\Omega^{D\; 2}_{23}-\Omega^{D \; 2}_{21}\right) v_1^2+\left(\Omega^{D\; 2}_{13}-\Omega^{D\; 2}_{11}\right) v_2^2\right) s_{13}c_{13}+\left(\Omega^{D}_{21} \Omega^{D}_{23} v_1^2+\Omega^{D}_{11} \Omega^{D}_{13} v_2^2\right) (s_{13}^{2} - c_{13}^{2}) ] \Bigg).
\end{align}

\section{Neutrino masses and rotation matrix}\label{neutrinomasses}

From Eq. (\ref{lagN}) 
we write the $9\times 9$ neutrino mass matrix in the  basis $\left(\begin{matrix}{\nu^{e,\mu,\tau}_{L}},\,\left(\nu^{e,\mu,\tau}_{R}\right)^{C},\,\left(N^{e,\mu,\tau}_{R}\right)^{C}\end{matrix}\right)$ as,

\begin{align}
\mathcal{M}_{\nu} &=\begin{pmatrix}
    0 & m_{D}^{T} & 0 \\
    m_{D} & 0 & M_{D}^{T} \\
    0 & M_{D} & M_{M}
    \end{pmatrix},     
\end{align}

where the block matrices are defined as,
\begin{align}
 m_{D}^{T}&=\frac{v_{2}}{\sqrt{2}}\begin{pmatrix}
    h_{2e}^{\nu e} & h_{2e}^{\nu \mu} & h_{2e}^{\nu \tau} \\
    h_{2\mu}^{\nu e} & h_{2\mu}^{\nu \mu} & h_{2\mu}^{\nu \tau} \\
    0 & 0 & 0
    \end{pmatrix}, &
     (M_{D})^{ij}&=\frac{v_{\chi}}{\sqrt{2}}h_{\chi i}^{\nu j}, &
     (M_{M})_{ij}&=\frac{1}{2}M_{N}^{ij}.
\end{align}

Neutrino masses are generated via inverse see-saw mechanism by assuming the hierarchy $M_{M}\ll m_{D} \ll M_{D} $. Block diagonalization is achieved by the rotation matrix $\mathbb{V}_{SS}$ given by,

\begin{align}
\mathbb{V}_{SS}\mathcal{M}_{\nu}\mathbb{V}_{SS}^{\dagger}&\approx\begin{pmatrix}
m_{light}&0\\
0&m_{heavy}
\end{pmatrix},  &
\mathbb{V}_{SS}&=\begin{pmatrix}
I&-\Theta_{\nu}\\
\Theta_{\nu}^{\dagger}&I
\end{pmatrix}, &
\Theta_{\nu}&=\begin{pmatrix}
m_{D}^{\dagger} & 0
\end{pmatrix}\begin{pmatrix}
0&M_{D}^{T}\\
M_{D}&M_{M}
\end{pmatrix}^{-1*},
\end{align}

\noindent
where $m_{light}=m_{D}^{T}(M_{D}^{T})^{-1}M_{M}(M_{D})^{-1}m_{D}$ is the $3\times3$ mass matrix containing the active neutrinos and $m_{heavy}$ contains the six heavy Majorana neutrino mass eigenstates, which reads,

\begin{align}
    m_{heavy}\approx\begin{pmatrix}0&M_{D}^{T}\\
M_{D}&M_{M}
\end{pmatrix}.
\end{align}

For simplicity, let's consider the case of $M_{D}$ being  diagonal and $M_{M}$ proportional to the identity.
\begin{align}\label{nuheavy}
M_{D} &= \frac{v_{\chi}}{\sqrt{2}} \left( \begin{matrix}
h_{N\chi e}	&	0	&	0	\\	0	&	h_{N\chi \mu}	&	0	\\	0	&	0	&	h_{N \chi \tau}
\end{matrix} \right), &
M_{M} &= \mu_{N} \mathbb{I}_{3\times 3} .  
\end{align}

Thus, light neutrino mass matrix takes the form,

\begin{equation}\label{mnu}
m_{\mathrm{light}} = \frac{\mu_{N} v_{2}^{2}}{{h_{N\chi e}}^{2}v_{\chi}^{2}}
\left( 
\begin{matrix}
	\left( h_{2e}^{\nu e}\right)^{2} + \left( h_{2\mu}^{\nu e} \right)^{2} \rho^{2} &
	{h_{2e}^{\nu e}}\,{h_{2e}^{\nu \mu}} + {h_{2\mu}^{\nu e}}\,{h_{2\mu}^{\nu \mu}}\rho^2 	&
	{h_{2e}^{\nu e}}\,{h_{2e}^{\nu \tau}}+ {h_{2\mu}^{\nu e}}\,{h_{2\mu}^{\nu \tau}}\rho^2 	\\
	{h_{2e}^{\nu e}}\,{h_{2e}^{\nu \mu}} + {h_{2\mu}^{\nu e}}\,{h_{2\mu}^{\nu \mu}}\rho^2	&	
	\left( h_{2e}^{\nu \mu} \right)^{2} + \left( h_{2\mu}^{\nu\mu} \right)^{2} \rho^{2}	&	
	{h_{2e}^{\nu \mu}}\,{h_{2e}^{\nu \tau}}+ {h_{2\mu}^{\nu \mu}}\,{h_{2\mu}^{\nu \tau}}\rho^2	\\
	{h_{2e}^{\nu e}}  \,{h_{2e}^{\nu \tau}}+ {h_{2\mu}^{\nu e}}  \,{h_{2\mu}^{\nu \tau}}\rho^2	&	
	{h_{2e}^{\nu \mu}}\,{h_{2e}^{\nu \tau}}+ {h_{2\mu}^{\nu \mu}}\,{h_{2\mu}^{\nu \tau}}\rho^2	&	
	\left( h_{2e}^{\nu \tau} \right)^{2} + \left( h_{2\mu}^{\nu \tau} \right)^{2} \rho^{2}
\end{matrix} \right),
\end{equation}

\noindent
where $\rho={h_{N\chi e}}/{h_{N\chi \mu}}$. $m_{\mathrm{light}}$ has rank $2$ so it contains a massless neutrino which is still allowed because experiments provide squared mass differences. Besides, we see that there is an overall factor which we assume to be the responsible of providing the mass energy scale. However, exotic neutrinos mass eigenstates, $\mathcal{N}^{k}$, $k=1,...,6.$, can be obtained easily from Eq. (\ref{nuheavy}) being the mass eigenvalues given by,
    \begin{align}
    m_{\mathcal{N}^{1}}&=\frac{1}{2}(\mu_{N}-\sqrt{\mu_{N}^{2}+2h_{N_{\chi e}}^{2}v_{\chi}^{ 2}}), & m_{\mathcal{N}^{4}}&=\frac{1}{2}(\mu_{N}+\sqrt{\mu_{N}^{2}+2h_{N_{\chi e}}^{2}v_{\chi}^{ 2}}), \\
    m_{\mathcal{N}^{2}}&=\frac{1}{2}(\mu_{N}-\sqrt{\mu_{N}^{2}+2h_{N_{\chi \mu}}^{2}v_{\chi}^{ 2}}), & m_{\mathcal{N}^{5}}&=\frac{1}{2}(\mu_{N}+\sqrt{\mu_{N}^{2}+2h_{N_{\chi \mu}}^{2}v_{\chi}^{ 2}}), \\
    m_{\mathcal{N}^{3}}&=\frac{1}{2}(\mu_{N}-\sqrt{\mu_{N}^{2}+2h_{N_{\chi \tau}}^{2}v_{\chi}^{ 2}}), & m_{\mathcal{N}^{6}}&=\frac{1}{2}(\mu_{N}+\sqrt{\mu_{N}^{2}+2h_{N_{\chi \tau}}^{2}v_{\chi}^{ 2}}). 
\end{align}
\noindent
Finally, the diagonal mass eigenstates are given by  $\mathcal{M}_{\nu}^{diag} = \mathcal{R}\mathcal{M}_{\nu}\mathcal{R}^{\dagger}$ where the rotation matrix is given by,

\begin{align}
\mathcal{R}&\approx \left( \begin{array}{c|cc}   
         V^{\nu}&  0 &   V^{\nu} m_{D}^{\dagger} M_{D}^{-1 *} \\\hline
       -\frac{i}{\sqrt{2}} M_{D}^{-1T} m_{D} &i\frac{1}{\sqrt{2}}\mathbb{I} & -i\frac{1}{\sqrt{2}}\mathbb{I}\\
       \frac{1}{\sqrt{2}}M_{D}^{-1T}m_{D} &\frac{1}{\sqrt{2}}\mathbb{I} &\frac{1}{\sqrt{2}}\mathbb{I}
    \end{array} \right),
\end{align}
\noindent
where $V^{\nu}$ is the rotation matrix for active neutrinos. Since the term $\mu_{N}$ is very small ($\mu_{N} \sim 2 m_{\mathcal{N}}^{2} \times 10^{-12}$ GeV) in comparison to $v_{\chi}$, it can be neglected making the first three exotic neutrino mass eigenvalues to be negative. which makes exotic neutrinos mass eigenstates nearly degenerate, so the $i$ factor in the second row arises to make all eigenvalues positive.

 \bibliographystyle{apsrev4-1}
 \bibliography{References.bib}

\end{document}